\documentclass[%
 reprint,
superscriptaddress,
preprintnumbers,
 amsmath,amssymb,
 aps, prl
]{revtex4-2}

\usepackage{graphicx}% Include figure files
\usepackage{dcolumn}% Align table columns on decimal point
\usepackage{bm}% bold math
\usepackage{hyperref}% add hypertext capabilities
\usepackage[mathlines]{lineno}% Enable numbering of text and display math
\usepackage{placeins}   % \FloatBarrier
\usepackage{booktabs}   % \toprule, \midrule, \bottomrule
\usepackage{adjustbox}  % for \adjustbox{max width=\textwidth}
\usepackage{xcolor}

\providecommand{\psipvis}[0]{\(\psi^{\prime}_{\text{vis}}\)}

\begin{document}

\preprint{FERMILAB-PUB-26-0707-PPD}

\title{First Measurement of the Superscaling Variable as a Function of Visible Kinematics in Charged-Current Quasi-Elastic Neutrino--Nucleus Scattering}
%\title{Measuring Nuclear Effects in Charged-Current Quasi-Elastic neutrino--nucleus Scattering with the Superscaling Variable}

%%%%%%%%%%%%%%%%%%%%%%%%%%%%%%%%%%%%%%%%%%%%%%%%%%%%%%%%%%%%%%%%%%%%%%%%%%% INSERT GLAUCUS AUTHOR LIST HERE
%% List of institution addresses, in command form.
\newcommand{\Rutgers}{Rutgers, The State University of New Jersey, Piscataway, New Jersey 08854, USA}
\newcommand{\Hampton}{Hampton University, Dept. of Physics, Hampton, VA 23668, USA}
\newcommand{\Dortmund}{Institute of Physics, Dortmund University, 44221, Germany }
\newcommand{\Otterbein}{Department of Physics, Otterbein University, 1 South Grove Street, Westerville, OH, 43081 USA}
\newcommand{\JMU}{James Madison University, Harrisonburg, Virginia 22807, USA}
\newcommand{\Florida}{University of Florida, Department of Physics, Gainesville, FL 32611}
\newcommand{\UCIrvine}{Department of Physics and Astronomy, University of California, Irvine, Irvine, California 92697-4575, USA}
\newcommand{\CBPF}{Centro Brasileiro de Pesquisas F\'{i}sicas, Rua Dr. Xavier Sigaud 150, Urca, Rio de Janeiro, Rio de Janeiro, 22290-180, Brazil}
\newcommand{\PUCP}{Secci\'{o}n F\'{i}sica, Departamento de Ciencias, Pontificia Universidad Cat\'{o}lica del Per\'{u}, Apartado 1761, Lima, Per\'{u}}
\newcommand{\INRM}{Institute for Nuclear Research of the Russian Academy of Sciences, 117312 Moscow, Russia}
\newcommand{\Jlab}{Jefferson Lab, 12000 Jefferson Avenue, Newport News, VA 23606, USA}
\newcommand{\Pittsburgh}{Department of Physics and Astronomy, University of Pittsburgh, Pittsburgh, Pennsylvania 15260, USA}
\newcommand{\Guanajuato}{Campus Le\'{o}n y Campus Guanajuato, Universidad de Guanajuato, Lascurain de Retana No. 5, Colonia Centro, Guanajuato 36000, Guanajuato M\'{e}xico.}
\newcommand{\Athens}{Department of Physics, University of Athens, GR-15771 Athens, Greece}
\newcommand{\Tufts}{Physics Department, Tufts University, Medford, Massachusetts 02155, USA}
\newcommand{\WM}{Department of Physics, William \& Mary, Williamsburg, Virginia 23187, USA}
\newcommand{\FNAL}{Fermi National Accelerator Laboratory, Batavia, Illinois 60510, USA}
\newcommand{\Purdue}{Department of Chemistry and Physics, Purdue University Calumet, Hammond, Indiana 46323, USA}
\newcommand{\MCLA}{Massachusetts College of Liberal Arts, 375 Church Street, North Adams, MA 01247}
\newcommand{\UMD}{Department of Physics, University of Minnesota -- Duluth, Duluth, Minnesota 55812, USA}
\newcommand{\Northwestern}{Northwestern University, Evanston, Illinois 60208}
\newcommand{\UNI}{Facultad de Ciencias F\'{i}sicas, Universidad Nacional Mayor de San Marcos, CP 15081, Lima, Per\'{u}}
\newcommand{\Rochester}{Department of Physics and Astronomy, University of Rochester, Rochester, New York 14627 USA}
\newcommand{\Austin}{Department of Physics, University of Texas, 1 University Station, Austin, Texas 78712, USA}
\newcommand{\USM}{Departamento de F\'{i}sica, Universidad T\'{e}cnica Federico Santa Mar\'{i}a, Avenida Espa\~{n}a 1680 Casilla 110-V, Valpara\'{i}so, Chile}
\newcommand{\Geneva}{University of Geneva, 1211 Geneva 4, Switzerland}
\newcommand{\Chicago}{Enrico Fermi Institute, University of Chicago, Chicago, IL 60637 USA}
\newcommand{\hired}{}
\newcommand{\OregonState}{Department of Physics, Oregon State University, Corvallis, Oregon 97331, USA}
\newcommand{\oxford}{Oxford University, Department of Physics, Oxford, OX1 3PJ United Kingdom}
\newcommand{\umiss}{University of Mississippi, Oxford, Mississippi 38677, USA}
\newcommand{\upenn}{Department of Physics and Astronomy, University of Pennsylvania, Philadelphia, PA 19104}
\newcommand{\AMU}{Department of Physics, Aligarh Muslim University, Aligarh, Uttar Pradesh 202002, India}
\newcommand{\wroclaw}{University of Wroclaw, plac Uniwersytecki 1, 50-137 Wroa\l{}aw, Poland}
\newcommand{\Mohali}{Department of Physical Sciences, IISER Mohali, Knowledge City, SAS Nagar, Mohali - 140306, Punjab, India}
\newcommand{\CINVESTAV}{Departamento de Fisica Col. San Pedro Zacatenco, 07360 Mexico, DF, Av. Instituto PolitÃ©cnico Nacional, Mexico}
\newcommand{\york}{York University, Department of Physics and Astronomy, Toronto, Ontario, M3J 1P3 Canada}
\newcommand{\ND}{Department of Physics and Astronomy, University of Notre Dame, Notre Dame, Indiana 46556, USA}
\newcommand{\ICL}{The Blackett Laboratory,  Imperial College London,  London SW7 2BW, United Kingdom}
\newcommand{\warwick}{Department of Physics, University of Warwick, Coventry, CV4 7AL, UK}
\newcommand{\qmul}{G O Jones Building, Queen Mary University of London, 327 Mile End Road, London E1 4NS, UK}
\newcommand{\LLNL}{Nuclear and Chemical Sciences Division, Lawrence Livermore National Laboratory, Livermore, CA 94550, USA}

\newcommand{\LorenzoGiannessiThanks}{now at Johannes Gutenberg-Universität Mainz Institut für Physik, Mainz, Germany}
\newcommand{\adrianThanks}{Now at Department of Physics, Drexel University, Philadelphia, Pennsylvania 19104, USA}
\newcommand{\lazazuetareyesThanks}{now at Syracuse University, Syracuse, NY 13244, USA}

% 35 total signatories.

\author{L.~Giannessi}\thanks{\LorenzoGiannessiThanks}  \affiliation{\Geneva}
\author{S.~Akhter}                        \affiliation{\AMU}
\author{Z.~Ahmad~Dar}                     \affiliation{\WM}  \affiliation{\AMU}
\author{N.S.~Alex}                        \affiliation{\Rochester}
\author{M.~Sajjad~Athar}                  \affiliation{\AMU}
\author{J.~Felix}                         \affiliation{\Guanajuato}
\author{L.~Fields}                        \affiliation{\ND}
\author{R.~Gran}                          \affiliation{\UMD}
\author{E.Granados}                       \affiliation{\Guanajuato}  \affiliation{\Guanajuato}
\author{D.A.~Harris}                      \affiliation{\york}  \affiliation{\FNAL}
\author{A.~Klustov\'{a}}                  \affiliation{\ICL}
\author{M.~Kordosky}                      \affiliation{\WM}
\author{D.~Last}                          \affiliation{\Rochester}  \affiliation{\upenn}
\author{A.~Lozano}\thanks{\adrianThanks}  \affiliation{\CBPF}
\author{S.~Manly}                         \affiliation{\Rochester}
\author{W.A.~Mann}                        \affiliation{\Tufts}
\author{K.S.~McFarland}                   \affiliation{\Rochester}
\author{M.~Mehmood}                       \affiliation{\york}
\author{O.~Moreno}                        \affiliation{\WM}  \affiliation{\Guanajuato}
\author{J.G.~Morf\'{i}n}                  \affiliation{\FNAL}
\author{J.K.~Nelson}                      \affiliation{\WM}
\author{G.N.~Perdue}                      \affiliation{\FNAL}  \affiliation{\Rochester}
\author{C.~Pernas}                        \affiliation{\WM}
\author{M.A.~Ram\'{i}rez}                 \affiliation{\upenn}  \affiliation{\Guanajuato}
\author{R.D.~Ransome}                     \affiliation{\Rutgers}
\author{N.~Roy}                           \affiliation{\york}
\author{D.~Ruterbories}                   \affiliation{\Rochester}
\author{F.~S\'{a}nchez}                       \affiliation{\Geneva}
\author{H.~Schellman}                     \affiliation{\OregonState}
\author{C.J.~Solano~Salinas}              \affiliation{\UNI}
\author{D.S.~Correia}                     \affiliation{\CBPF}
\author{V.S.~Syrotenko}                   \affiliation{\Tufts}
\author{N.H.~Vaughan}                     \affiliation{\OregonState}
\author{A.V.~Waldron}                     \affiliation{\qmul}  \affiliation{\ICL}
\author{L.~Zazueta}\thanks{\lazazuetareyesThanks}  \affiliation{\WM}

\collaboration{The MINER$\nu$A Collaboration}
\noaffiliation
\date{\today}

%%%%%%%%%%%%%%%%%%%%%%%%%%%%%%%%%%%%%%%%%%%%%%%%%%%%%%%%%%%%%%%%%%%%%%%%%%%

\begin{abstract}
We report the first measurement of the superscaling variable $\psi^{\prime}_{\rm vis}$ in neutrino--nucleus scattering, using charged-current quasielastic-like $\nu_\mu$-Carbon interactions collected by the MINERvA experiment in the medium-energy NuMI beam. Originally developed to characterize phenomena of scaling with respect to atomic number and momentum transfer in quasielastic electron-nucleus scattering, $\psi^{\prime}$ is reconstructed here from visible final-state kinematics ($\psi^{\prime}_{\rm vis}$). $\psi^{\prime}_{\rm vis}$ is used here as an externally calibrated variable whose peak position is sensitive to the invisible nuclear missing energy. We measure $\psi^{\prime}_{\rm vis}$ as functions of the hadronic recoil $\Sigma T_p$ and muon transverse momentum $p_{\text{T}\mu}$, comparing background-subtracted data to five neutrino--nucleus interaction models via a forward-folding technique. The predicted $\psi^{\prime}_{\rm vis}$ diverges most strongly among model predictions at low $\Sigma T_p$ and $p_{\text{T}\mu}$, where agreement with data is also poorest, exposing which models fail to capture the phase-space dependence of the missing energy and therefore the neutrino energy as extrapolated from visible kinematics. Among tested models, we find that the Relativistic Mean Field approach predicts more accurately the $\psi^{\prime}_{\rm vis}$ peak position.
%These results provide a new way of benchmarking neutrino--nucleus interaction models through their modelling of the relationship between the observable energy deposits in the detector and the energy of the interacting neutrino, underpinning current and future long-baseline oscillation measurements.
\end{abstract}

%Use showkeys class option if keyword
\maketitle

% \section{Introduction}\label{sec:introduction}
% \input{introduction}

Measurements of neutrino oscillations at long-baseline experiments~\cite{T2K,NOVA,HK,DUNE} rely on neutrino--nucleus interaction models to relate the observed final-state kinematics to the incoming neutrino energy. Uncertainties in the interaction model may limit the sensitivity to oscillation parameters in next-generation experiments. Measuring differential neutrino--nucleus cross-sections with respect to leptonic and hadronic final-state kinematics~\cite{Lu:2015tcr,Lu:2015hua,Furmanski:2016wqo,MINERvA:2018hba,T2K:2018rnz,MicroBooNE:2023krv} may reveal mismodeled physics ingredients, and therefore help improving interaction models.

% Measuring neutrino--nucleus cross-sections and benchmarking interaction models has a twofold importance. Differential cross-sections with respect to final-state kinematics provide phenomenological inputs for models and constrain their parameters. Variables combining leptonic and hadronic final-state kinematics are able to identify nuclear effects and isolate interaction channels~\cite{Lu:2015tcr,Lu:2015hua,Furmanski:2016wqo,MINERvA:2018hba,T2K:2018rnz,MicroBooNE:2023krv}, and comparing model to lepton-hadron correlation data reveals mismodeled physics ingredients. 

% Moreover, variables that probe the relation between neutrino energy and observable final-state kinematics are especially important for oscillation experiments.

The superscaling variable $\psi^\prime$ was established in the study of \textit{superscaling} in quasielastic (QE) electron-nucleus scattering~\cite{donnelly_sick_1990,donnelly_sick_1998,donnelly_sick_1999,donnelly_sick_2002}, where it underpins phenomenological nuclear models exported to the neutrino sector~\cite{Amaro_scaling_2005,Caballero_scaling_2006,SUSA_review}, and has recently been shown to be accessible in semi-inclusive neutrino--nucleus scattering~\cite{Douqa_2024}. Rather than testing whether scaling holds in neutrino--nucleus scattering, we use $\psi^\prime$ as an externally calibrated observable to expose how interaction models share the neutrino energy between the visible final state and the invisible nuclear missing energy. We report a measurement of $\psi^\prime_{\rm vis}$ in neutrino--carbon charged-current interactions with no pions in the final state (quasielastic-like) in MINERvA~\cite{MINERVA}. The event rate in this variable is compared to several interaction models via a forward-folding approach in which truth-level predictions are passed through MINERvA's beam and detector simulations to produce reconstructed kinematics directly comparable to data. 
% In the context of neutrino--nucleus scattering, $\psi^\prime$ carries important information linking the reconstructed neutrino energy with the visible final state kinematics.  
% After introducing the neutrino-scattering superscaling variable $\psi^\prime_{\text{vis}}$ and its relation to neutrino energy reconstruction, we describe the apparatus and measurement methods, compare model predictions with data, and discuss results.

% \section{Superscaling variable and neutrino energy reconstruction}\label{sec:superscaling}
% \input{superscaling_short}

Scaling occurs when the lepton--nucleus cross section ($\sigma_{\ell A}$) can be expressed as the lepton--nucleon cross section ($\sigma_{\ell N}$) multiplied by a scaling function $f(z)$ such that $\sigma_{\ell A} = \sigma_{\ell N}\cdot f(z)$, where the scaling variable $z=z(\omega,|\vec q|)$ is an appropriate combination of the energy transfer, $\omega$, and three-momentum transfer, $|\vec{q}|$. The observation of scaling indicates that the probe interacts incoherently with a single bound nucleon rather than with the composite nucleus~\cite{PhysRevC.69.028501,Kawazoe,WEST1975,donnelly_sick_1990}. When the same scaling function is also independent of the target nucleus, scaling of the second kind, or $A$-scaling, is realized.  Simultaneous occurrence of both kinds is termed
\emph{superscaling}~\cite{donnelly_sick_1998,donnelly_sick_1999,donnelly_sick_2002}.

One combination of energy and momentum transfer is the dimensionless variable $\psi'$ introduced in electron-nucleus scattering~\cite{donnelly_sick_1998,donnelly_sick_1999}. Defining $\lambda=\omega/2m_{N}$, $\kappa=|\vec q|/2m_{N}$, where $m_{N}$ represents the target nucleon mass, and $\tau=\kappa^{2}-\lambda^{2}$, $\psi'$ is constructed as 
\begin{equation}
\psi^{\prime} \;=\; \frac{1}{\sqrt{\xi_F}}\,
\frac{\lambda^{\prime}-\tau^{\prime}}
{\sqrt{(1+\lambda^{\prime})\tau^{\prime}
+\kappa\sqrt{\tau^{\prime}(\tau^{\prime}+1)}}},
\label{eq:psiprime}
\end{equation}
where the shifted 
$\lambda^{\prime}=\lambda-E_{\rm shift}/(2m_{N})$ and
$\tau^{\prime}=\kappa^{2}-\lambda^{\prime 2}$ correct for the average energy required to extract the nucleon from the nuclear potential, and the normalizing factor $\xi_{F}=\sqrt{1+(k_{F}/m_{N})^{2}}-1$ depends only on the Fermi momentum $k_F$.
% One combination of energy and momentum transfer is the dimensionless variable $\psi'$ introduced in electron-nucleus scattering~\cite{donnelly_sick_1998,donnelly_sick_1999},
% \begin{equation}
% \psi^{\prime} \;=\; \frac{1}{\sqrt{\xi_F}}\,
% \frac{\lambda^{\prime}-\tau^{\prime}}
% {\sqrt{(1+\lambda^{\prime})\tau^{\prime}
% +\kappa\sqrt{\tau^{\prime}(\tau^{\prime}+1)}}},
% \label{eq:psiprime}
% \end{equation}
% where $\lambda=\omega/2m_{N}$, $\kappa=|\vec q|/2m_{N}$,
% $\tau=\kappa^{2}-\lambda^{2}$, $\xi_{F}=\sqrt{1+(k_{F}/m_{N})^{2}}-1$, and where the shifted 
% $\lambda^{\prime}=\lambda-E_{\rm shift}/(2m_{N})$ and
% $\tau^{\prime}=\kappa^{2}-\lambda^{\prime 2}$ correct for the average energy required to extract the nucleon from the nuclear potential. 
The Fermi momentum ($k_{F}$) and the energy shift ($E_{\rm shift}$) parameters entirely encode the dependence of $\psi^{\prime}$ on the target nucleus, and are
extracted from electron-scattering data and tabulated for a wide range of nuclei~\cite{donnelly_sick_2002}.
%The two parameters play distinct roles in shaping the QE distribution: 
While $E_{\rm shift}$ controls the \emph{position} of the QE peak by displacing it from the elastic $\lambda=\tau$ to the QE $\lambda^{\prime}=\tau^{\prime}$, $k_{F}$ controls the \emph{width} of the distribution. $E_{\rm shift}$ and $k_{F}$ are linked to the nuclear structure, therefore negligible difference is expected between the neutrino and the electron case; for Carbon we fix $E_{\rm shift}=20$~MeV, $k_{F}=228$~MeV/c as observed in electron scattering~\cite{donnelly_sick_2002}.
The crucial property of $\psi^{\prime}$ is kinematical: the $\psi^\prime$ distribution from a sample of QE events is expected to peak at $\psi^{\prime}=0$, with deviations expected in regions of the phase space where the neutrino energy reconstruction is not accurate, or the parameter $E_{\rm shift}$ no longer represents the average removal energy. 
% Superscaling analyses in electron scattering has informed modern phenomenological neutrino--nucleus interaction models, this work is the first measurement of $\psi^\prime$ in neutrino scattering.

The neutrino measurement of $\psi^{\prime}$ differs from the electron case in that the incoming lepton energy is not known {\em a priori}, and the
four-momentum transfer $(\omega,\vec q)$ must be reconstructed from the
final state. Additionally, a correction due to the charge-exchange in the interaction, proportional to the square of the neutron-proton mass difference, must be applied~\cite{amziane26}, although this is negligible at MINERvA energies.
%Second, the charge-exchange nature of the charged-current vertex modifies the elastic condition by $|\vec q|^{2}\rightarrow |\vec q|^{2}+\Delta M^{2}$, with $\Delta M^{2}=m_{n}^{2}-m_{p}^{2}\simeq 2.4\times 10^{-3}~\mathrm{GeV}^{2}$. This correction, however, is negligible compared to the typical $|\vec q|^{2}\sim 0.4$--$0.5~\mathrm{GeV}^{2}$ at MINERvA.
Typically, only part of the energy deposited by a neutrino leaves a detectable signal: $E_\nu = E_{\rm vis} + E_{\rm miss}$. $E_{\rm vis}$ is the energy deposited in the detector by final state particles, and $E_{\rm miss}$ is the undetected nuclear missing energy.
%The neutrino energy estimator used in Cherenkov detectors, $E_{\nu}^{\rm QE}$, is the inversion of the condition $\psi^{\prime}=0$, assuming the lepton kinematics that would follow from a strictly QE process, and therefore inherits a bias whenever the true event lies away from the QE peak. 
In this work we adopt the ``calorimetric'' neutrino energy estimator, typically used in detectors with the ability to measure the hadronic final state~\cite{MINERVA,T2K,ND280upgrade,NOvA:2020rbg}:
\begin{equation}
E_{\nu}^{\mathrm{RE}} \;=\; E_{\mu} + \sum T_{p} + m_{p} - m_{n} + S_{\mathrm{RE}},
\label{eq:enure}
\end{equation}
where $m_p$ and $m_n$ are respectively the proton and neutron masses, $S_{\mathrm{RE}}$ is a constant representing the average nuclear missing energy~\cite{Douqa_2024}, and $\sum T_{p}$ is the sum of the kinetic energy of all the final state protons. The reconstructed components of the momentum transfer are therefore defined as:
\begin{align}
    \omega_{RE} &= \sum T_{p} + m_{p} - m_{n} + S_{\mathrm{RE}} ,\\
    \vec q^{T}_{RE} &=  -\vec p_{\text{T}\mu},\\
    q^L_{RE} &= \sum T_{p} + m_{p} - m_{n} + S_{\mathrm{RE}} + (E_{\mu}-|\vec p_{\text{L}\mu}|).
    \label{eq:momentum_transfers}
\end{align}
where $\vec q^{T}_{RE}$ and $q^{L}_{RE}$ are, respectively, the transverse and longitudinal momentum transfers estimated with the neutrino energy reconstruction formula~\ref{eq:enure}, from which the \emph{visible} superscaling variable \psipvis{} is measured.

%\paragraph{Quasielastic condition}
% A sample of QE lepton-nucleus events will have a $\psi^{\prime}_{\rm vis}$ distribution peaked at 0 with the proper choice of $E_{\rm shift}$ and $S_{RE}$. We fix $E_{\rm shift}=20$~MeV from electron-nucleus scattering data, and $S_{RE}=28$~MeV, roughly the average amount of nuclear missing energy in charged-current quasielastic (CCQE) neutrino--nucleus scattering~\cite{KDAR}. By selecting specific visible final state kinematics, we select a sample of events for which the average missing energy is not necessarily equal to the average $S_{RE}$, causing displacement of the $\psi^{\prime}_{\rm vis}$ peak from 0. 
% A model that reproduces this displacement in all regions of the phase space correctly models the average nuclear missing energy across final state kinematics, therefore yielding an accurate estimation of the neutrino energy from the visible hadronic and leptonic final state.
A sample of QE-enriched lepton-nucleus events has a $\psi^{\prime}_{\rm vis}$ distribution peaked at 0 if the reconstructed neutrino energy has negligible bias, that is if $S_{RE}$ is properly chosen. We fix $S_{RE}=28$~MeV, roughly the average nuclear missing energy in charged-current quasielastic (CCQE) neutrino--nucleus scattering~\cite{KDAR}. Selecting specific visible final-state kinematics selects a sample whose average missing energy need not equal $S_{RE}$, displacing the $\psi^{\prime}_{\rm vis}$ peak from 0. This displacement is a manifestation of the kinematic mismatch between the QE-on-static-nucleon assumption and the actual event kinematics: the nuclear missing energy in that region of phase space departs from the average value assumed in the reconstruction, hence a bias in the reconstructed neutrino energy is expected. 
%(in which case truly $\psi^{\prime}\neq 0$). 
A model that reproduces this displacement across the whole phase space therefore captures the phase-space dependence of the nuclear missing energy correctly, yielding an accurate estimation of the neutrino energy from the visible hadronic and leptonic final state and prediction of the expected neutrino energy bias. 
% Therefore, measuring $\psi^{\prime}_{\rm vis}$ offers a handle to understand model shortcomings in relating neutrino energy to final state kinematics, an important source of systematic uncertainty in the determination of the squared mass splitting $\Delta m^2_{23}$~\cite{dolan2026characterisingrolefinalstate}. This concern sharpens for next-generation, statistics-rich experiments such as DUNE~\cite{DUNE} and Hyper-Kamiokande, where comparable mismodeling of the visible-to-true energy relationship has been shown to bias oscillation-parameter extraction beyond the projected precision of these measurements.
The measurement tests nuclear-model ingredients relevant to calorimetric and QE-based energy reconstruction used in long-baseline oscillation analyses as the T2K-NOvA joint fit~\cite{abubakarJointNeutrinoOscillation2025b}.

A further strength of $\psi^{\prime}_{\rm vis}$ is that it separates the QE component from non-QE backgrounds: the QE-like events that are not QE populate $\psi^{\prime}_{\rm vis}$ in characteristic, well-separated regions: multi-nucleon knockout (2p2h), resonant and deep-inelastic scattering (DIS) interactions populate the high-$\psi^{\prime}_{\rm vis}$ side of the distribution. The central peak is dominated by the QE component, while these backgrounds form a broad tail that barely encroaches on $\psi^{\prime}_{\rm vis}\!\sim\!0$~\cite{Douqa_2024}, which allows measurement of $\psi^{\prime}_{\rm vis}$ on reasonably pure QE-like samples. On the other hand, genuine QE events where proton energy is lost due to nuclear re-scattering populate the negative side of the distribution.

For the measurement presented in this Letter we analyzed muon neutrino interactions on hydrocarbon collected by the MINERvA~\cite{MINERVA} experiment in the medium-energy (ME) NuMI beam~\cite{NUMI}, with an exposure of 1.06$\times10^{21}$ protons on target (POT). The ME NuMI beam is a wide-band, high-intensity neutrino beam produced at Fermi National Accelerator Laboratory with peak energy $\sim6$~GeV. The MINERvA detector, described in detail in Ref.~\cite{MINERVA}, is composed of an 8-tonnes scintillating tracker surrounded by semi-active lead- and steel-scintillator calorimeters. The MINOS~\cite{MINOS} near detector, a magnetic spectrometer, directly downstream of the MINERvA detector, measures the momentum and charge of escaping muons. Muon transverse momentum $p_{\text{T}\mu}$ is determined from the muon track curvature and range, and the measurement of $\Sigma T_p$ uses the calorimetric energy deposited in the scintillator bars not associated with the muon track.
Scintillator quenching and energy deposits mistakenly attributed to protons bias the reconstruction of $\Sigma T_p$ from detector signals, with the bias growing as the measured $\Sigma T_p$ grows.
To reduce this bias, the reference neutrino--nucleus interaction model,  \texttt{MnvTune4.4.1}~\cite{collaborationSimultaneousMeasurementProton2022}, and  the \textsc{GEANT4}~\cite{geant1,geant2,geant3} simulation of the detector are used to calibrate $\Sigma T_p$ against the observed energy deposit.
Systematic uncertainties cover plausible variations of the underlying model relevant for $\Sigma T_p$ calibration; the dominating systematic uncertainties are caused by the simulation's ability to model neutron secondary interactions. Detector response uncertainties are constrained by test beam analyses using incoming protons, pions and electrons.~\cite{MINERvA:2015yej} 

\begin{figure*}[tp]
    \centering
    \includegraphics[width=0.9\linewidth]{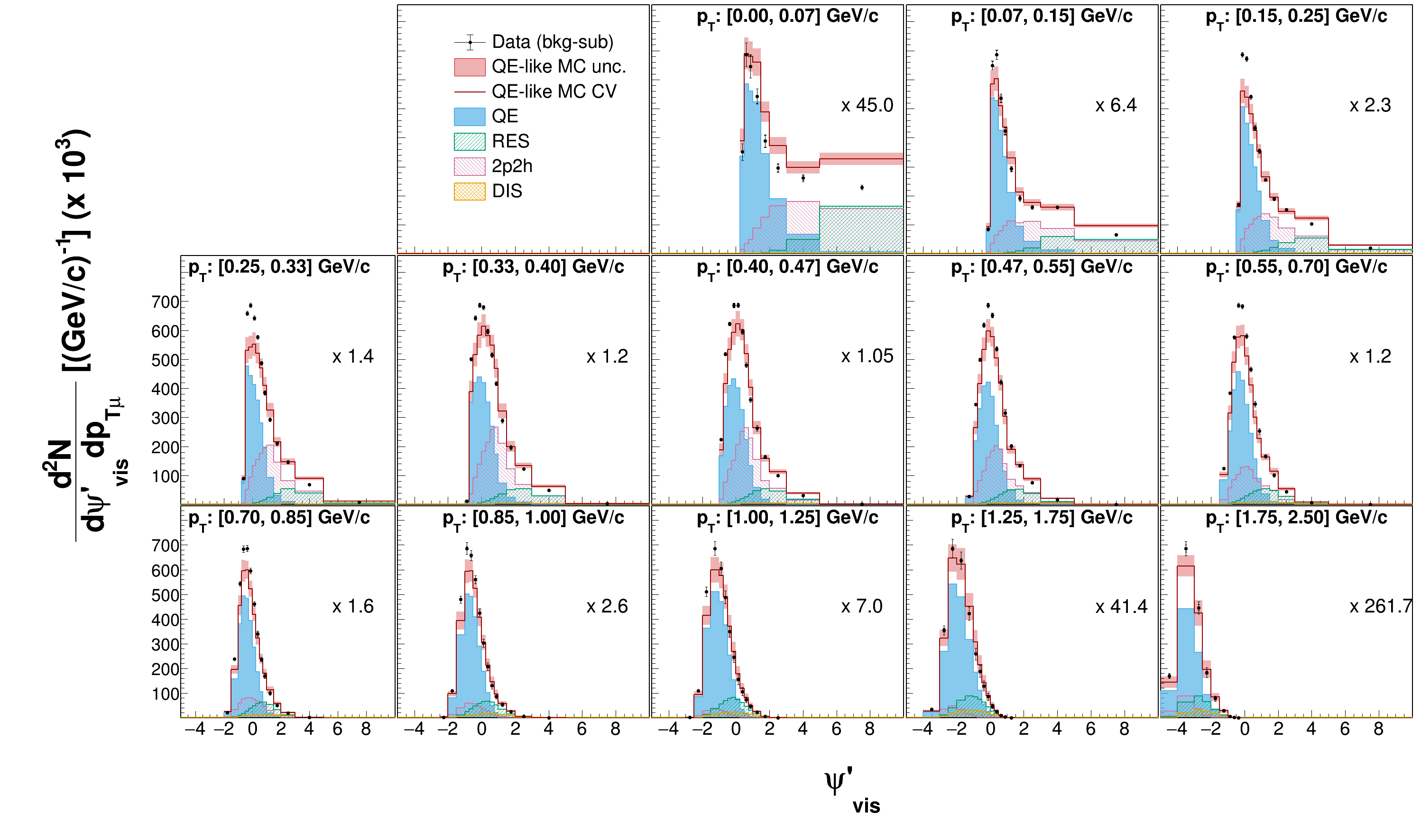}
    \caption{CCQE-like prediction of the reference model \texttt{MnvTune4.4.1} and the background-subtracted data of the reconstructed \psipvis{} distribution in bins of the muon transverse momentum $p_{\text{T}\mu}$. Error bars on both MC and Data histograms represent the systematic and statistical errors, summed in quadrature. At low $p_{\text{T}\mu}$, even though the irreducible background is relevant, the peak is still mostly populated by true QE events, with non-QE contaminations populating the high-\psipvis{} tail.}
    \label{fig:MnvTune_Pt}
\end{figure*}
\begin{figure*}[tp]
    \centering
    \includegraphics[width=0.9\linewidth]{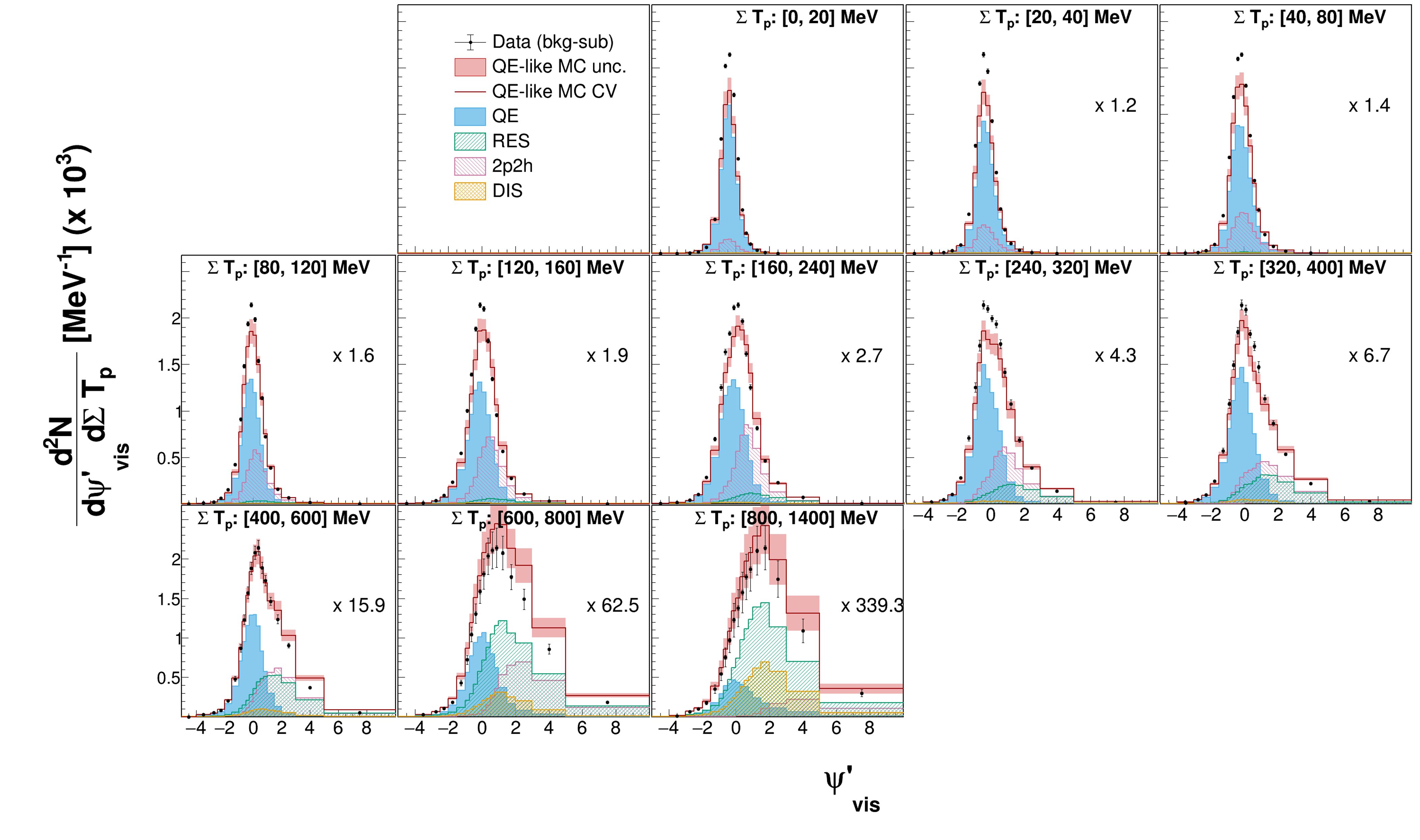}
    \caption{CCQE-like prediction of the reference model \texttt{MnvTune4.4.1} and the background-subtracted data of the reconstructed \psipvis{} distribution in bins of the hadronic recoil $\Sigma T_p$. Error bars on both MC and Data histograms represent the systematic and statistical errors summed in quadrature. Although the background contaminations populate different regions of \psipvis{} with respect to the QE sample, large background affects the peak position at $\Sigma T_p$ larger than $\sim240$~MeV.}
    \label{fig:MnvTune_SumTp}
\end{figure*}

The $\psi^{\prime}_{\rm vis}$ quantity is computed in each event from the final state lepton and hadron kinematics using Eqs.~\ref{eq:psiprime} and~\ref{eq:enure}.
In this analysis, we present a measurement of $\psi^{\prime}_{\rm vis}$ as a function of $\Sigma T_p$ and of $p_{\text{T}\mu}$ independently to benchmark models throughout the kinematic phase space.

%The analysis scrutinizes charged-current quasielastic (CCQE) scattering of muon neutrinos on carbon. 
The topological event selection reduces pion backgrounds to CCQE-like events based on the number of identified Michel electrons from the $\pi^+\to\mu^+\to e^+$ decay chain and isolated clusters from $\pi^0\to\gamma\gamma$ in the fiducial volume, yielding 1.45~million neutrino events in the CCQE-like signal sample.   
Three background-enriched samples are used to estimate the number of background events in signal sample as a function of $p_{\text{T}\mu}$ and $\sum T_p$~\cite{BVLS}. The resulting data-aided background estimation allows to obtain a background-subtracted sample in each bin of $p_{\text{T}\mu}$, $\sum T_p$,  and $\psi^{\prime}_{\rm vis}$.   This constrained background subtraction is described in detail in Ref.~\cite{collaborationSimultaneousMeasurementProton2022,PhysRevD.99.012004}. 
% :
% \begin{align}
%     N_i^{\text{MC,tuned}}=\sum_{j}N_{i}^{j}(p_{\text{T}\mu}, \Sigma T_p, \psi^\prime) \\ \times w^{j}(p_{\text{T}\mu}, \Sigma T_p),
% \end{align}
% where $j$ runs over the three dominant background processes selected by the sidebands.
% This is used to obtain the background-subtracted data in each bin of the ($p_{\text{T}\mu}$, $\sum T_p$, $\psi^{\prime}_{\rm vis}$) space, compared with the Monte Carlo model prediction of the reconstructed signal process in the same bins.

We compare background-subtracted data to flux-averaged generator predictions of the CCQE-like signal convoluted with MINERvA detector smearing due to finite resolution. This method is sometimes referred to as \textit{forward-folding}~\cite{Koch_2019}, and it avoids the model dependencies inherent to the unfolding procedure required to obtain true-level information from reconstructed quantities. Beyond the unavoidable reliance on \textsc{GEANT4} for detector modelling, model dependence enters this analysis only through the $\Sigma T_p$ calibration described above. 

Systematic uncertainties are estimated using the reference interaction model, by evaluating the impact of plausible detector, flux and interaction model variations on reconstructed event rates, as detailed in the End Matter. 

% \section{Interaction models}\label{sec:models}
% \input{models_short}

% \section{Results and discussion}\label{sec:results}
% \input{results}

We present results of the reconstructed \psipvis{} after background subtraction in bins of $p_{\text{T}\mu}$ (Fig.~\ref{fig:MnvTune_Pt}) and $\Sigma T_p$ (Fig.~\ref{fig:MnvTune_SumTp}) to study dependence on the lepton and hadron kinematics, respectively. The reference model \texttt{MnvTune4.4.1} prediction is overlaid for comparison. Non-QE but CCQE-like events are separated into 2p2h interactions, pion production from baryon resonances (RES), and deep inelastic scattering where any produced hadrons that are not nucleons have been absorbed in the nucleus (DIS). The QE contribution to the CCQE-like signal dominates the \psipvis{} peak position except at large hadronic recoil. By contrast, the low hadronic recoil region ($\Sigma T_p<80$~MeV) in Fig.~\ref{fig:MnvTune_SumTp} is predicted to be almost entirely QE and thus allows benchmarking of the QE model. The peak position and bulk of the \psipvis{} distribution shift from large to small values with increasing $p_{\text{T}\mu}$, and from near zero to large positive values with increasing $\Sigma T_p$. Events with large hadronic recoil -- in which a large fraction of the neutrino energy is transferred to the hadronic system -- are identified by a strongly positive \psipvis{} peak. On the other hand, events with very high $p_{\text{T}\mu}$ tend toward negative \psipvis{} due to detector smearing which on average migrates events to larger $p_{\text{T}\mu}$ because of the steeply falling true rate with increasing $p_{\text{T}\mu}$. These behaviors are broadly reproduced by all tested models, albeit with notable differences in specific bins. The reference model reproduces most features of the data, though it underestimates the cross section at $0.07~\text{GeV/c}<p_{\text{T}\mu}<0.4~\text{GeV/c}$~\cite{MINERvA:2015ydy} and for $\Sigma T_p<$120~MeV. 

\begin{figure}[tp]
    \centering
    \includegraphics[width=1\linewidth]{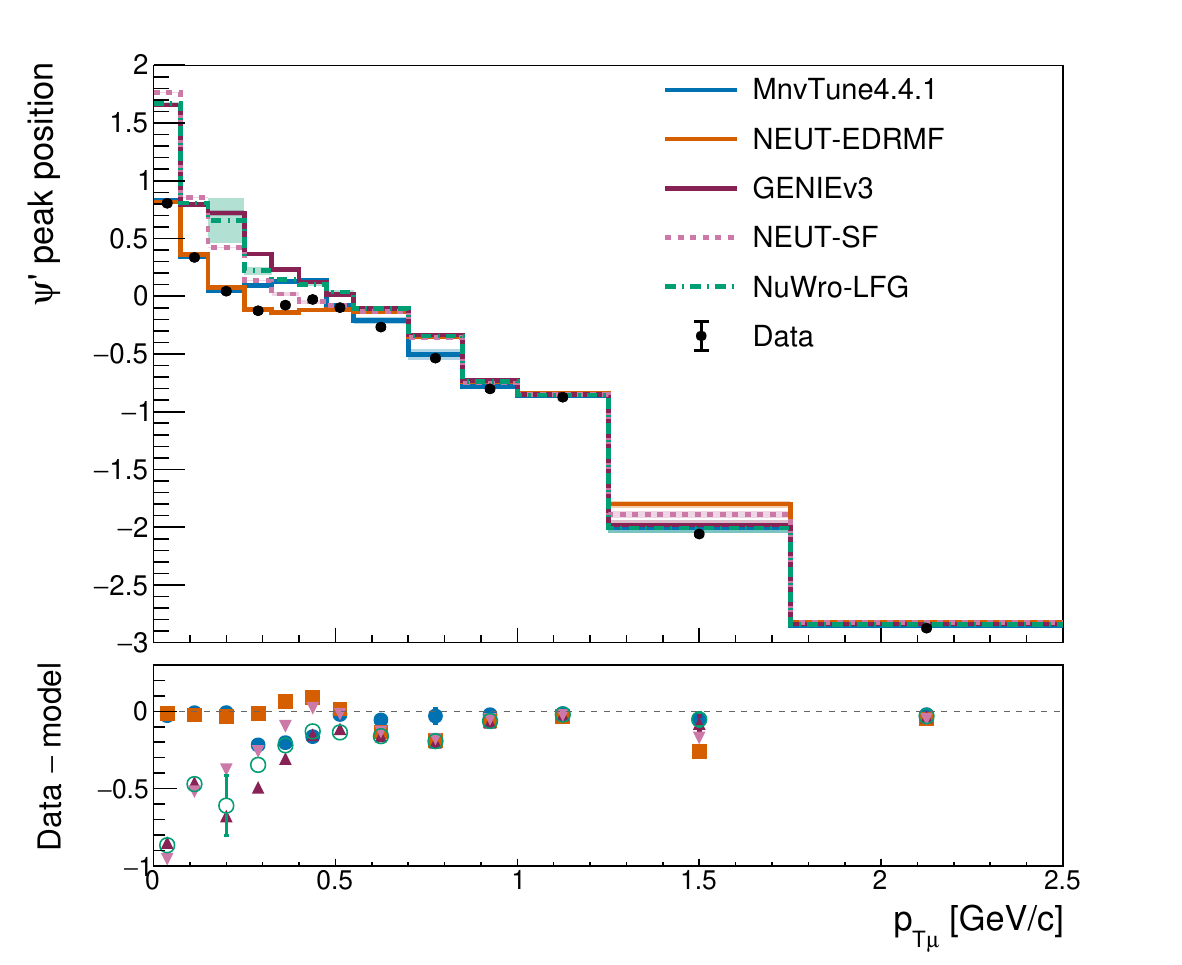}
    \caption{\psipvis{} peak position as a function of $p_{\text{T}\mu}$. Error bars represent the systematic uncertainties impact on the determination of the peak of the MC signal and background-subtracted data distributions. Tested models overpredict the \psipvis{} peak position, with larger disagreement with data at low $p_{\text{T}\mu}$.}
    \label{fig:peaks_pt}
\end{figure}

\begin{figure}[tp]
    \centering
    \includegraphics[width=1\linewidth]{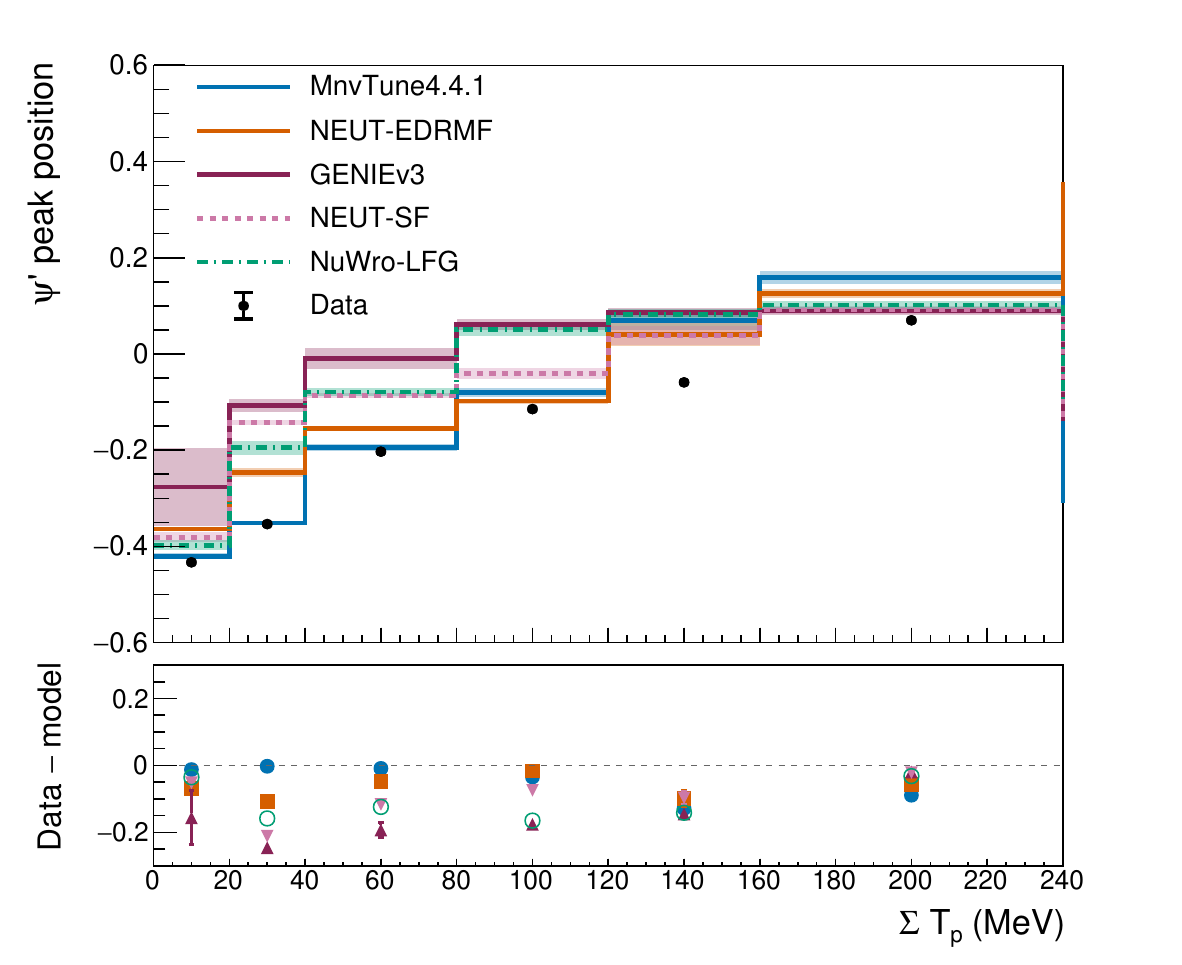}
    \caption{\psipvis{} peak position as a function of $\Sigma T_{p}$. Error bars computed as in Fig.~\ref{fig:peaks_pt}. All models under test overpredict the \psipvis{} peak position across the $\Sigma T_{p}$ range.}
    \label{fig:peaks_sumtp}
\end{figure}

We analyze the measured \psipvis{} peak position and compare it to several model predictions (Figs.~\ref{fig:peaks_pt} and~\ref{fig:peaks_sumtp}). Details on the interaction models and generators tested can be found in the End Matter. The peak position is estimated for data and Monte Carlo (MC) simulation using kernel density estimation~\cite{wand1995kernel}, which provides a robust estimator and uncertainty quantification. Systematic uncertainties are derived by estimating the peak with plausible model variations (detailed in the End Matter), while statistical uncertainties come from Poisson fluctuations of the \psipvis{} histograms. Full distributions, from which the peak positions are estimated, are shown in the Supplemental Material. %(Figs.~\ref{fig:allmodels_Pt} and~\ref{fig:allmodels_SumTp}) 

Model predictions of the \psipvis{} peak position differ the most at low $p_{\text{T}\mu}$, where agreement with data is also poorest. Figure~\ref{fig:peaks_pt} shows the \psipvis{} peak position as a function of $p_{\text{T}\mu}$. Since the energy shift from nuclear re-scattering carries little $p_{\text{T}\mu}$ dependence, it contributes only a constant offset across $p_{\text{T}\mu}$ bins, so the observed trend is sensitive primarily to the removal energy of the bound target neutron.
% The \psipvis{} peak position as a function of $p_{\text{T}\mu}$ shown in Fig.~\ref{fig:peaks_pt} integrates over the hadronic recoil and therefore the visible energy shift due to nuclear re-scattering is roughly average of the full phase space, which should be sensitive primarily to the removal energy of the bound target neutron.
For $p_{\text{T}\mu}<$0.5~GeV/c, models predict a higher \psipvis{} peak position with the exception of \texttt{MnvTune4.4.1} and \texttt{NEUT-EDRMF}~\cite{jake_EDRMF,NEUT}. Both models show better neutrino energy reconstruction at low momentum transfer with respect to the spectral function~\cite{BENHAR1994493} and local Fermi gas~\cite{valverdeQuasielasticNeutrinonucleusReactions2006,bourguilleInclusiveExclusiveNeutrinonucleus2021} approaches -- \texttt{MnvTune4.4.1} through its tuning to low-energy neutrino data, and \texttt{NEUT-EDRMF} through a sounder theoretical treatment of the nuclear state~\cite{benchmarking_alexis,alexis_microboone}.

Figure~\ref{fig:peaks_sumtp} reports the estimated \psipvis{} in the region $0~\text{MeV}<\Sigma T_p<~240~\text{MeV}$, where the QE component dominates enough to determine the QE-like peak position. Due to lower QE purity, the peak position estimation is less reliable in higher $\Sigma T_p$ (as seen in Fig.~\ref{fig:MnvTune_SumTp}), which are therefore excluded from the peak analysis. At low $\Sigma T_p$, the \psipvis{} peak position is governed primarily by the nucleon removal energy and by the impact of nuclear re-scattering. Overall, the tuned MINERvA model shows the best agreement with data. The \texttt{NEUT-EDRMF} model exhibits a less pronounced overprediction of the peak, consistent with a smaller underestimation of the nuclear removal energy relative to the other models. The \texttt{GENIE-v3} model shows the largest \psipvis{} overestimation, ascribed to overprediction of the visible hadronic energy, previously reported in~\cite{ascencio}.

\begin{figure}[tp]
    \centering
    \includegraphics[width=1\linewidth]{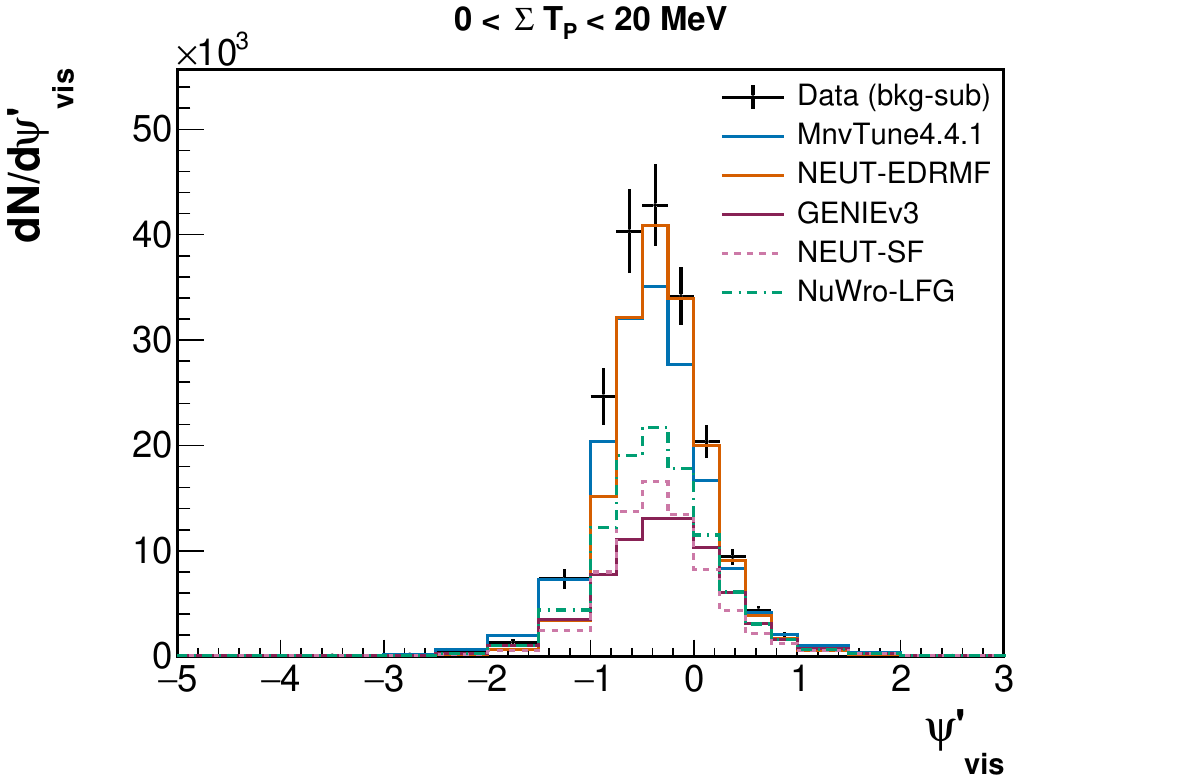}
    \caption{Reconstructed \psipvis{} for $\Sigma T_p<20$~MeV. Error bars on the data histogram include data statistical uncertainty and systematic model uncertainties. Underprediction of the cross-section by models \texttt{GENIE-v3} \texttt{NEUT-SF} and \texttt{NuWro-LFG} is substantial.}
    \label{fig:first_sumtp_bin}
\end{figure}

The first bin ($\Sigma T_p < 20$~MeV) is the only bin where \texttt{GENIE-v3}, \texttt{NuWro-LFG} and \texttt{NEUT-SF} reproduce the \psipvis{} peak position marginally better than \texttt{NEUT-EDRMF}. At such low proton kinetic energy the modelling of Pauli blocking becomes decisive, and only \texttt{NEUT-EDRMF} derives it self-consistently from quantum mechanics rather than through a simple momentum cut. Accordingly, it is the model yielding the best event rate prediction in this bin (Fig.~\ref{fig:first_sumtp_bin}). The other three substantially underpredict the event rate, their low-$\Sigma T_p$ entries originating from higher initial momenta degraded by nuclear re-scattering, and their marginally better peak position reflects only the kinematics of re-scattered protons.

% The first $\Sigma T_p$ bin, where the peak position heavily depends on the impact of nuclear re-scattering. Figure~\ref{fig:first_sumtp_bin} shows the \psipvis{} distribution for $\Sigma T_p\,<\,20$~MeV, where \texttt{NEUT-EDRMF} yields the prediction closest to data. As the only model deriving Pauli blocking self-consistently from quantum mechanics rather than via simple momentum cuts, it suggests first-principles treatments are essential for precision in low proton kinetic energy regions. Conversely, \texttt{GENIE-v3}, \texttt{NuWro-LFG}, and \texttt{NEUT-SF} predict almost no events with such low primary $\Sigma T_p$; their few entries originate from higher initial momenta reduced by nuclear re-scattering.

This work presents the first measurement of differential cross sections in the visible scaling variable \psipvis{} for neutrino quasielastic scattering, providing a new observable sensitive to the nuclear missing energy of the bound target neutron. From this measurement we draw three main conclusions. First, the \psipvis{} peak position, studied as a function of $p_{\text{T}\mu}$ (Fig.~\ref{fig:peaks_pt}) and of $\Sigma T_p$ (Fig.~\ref{fig:peaks_sumtp}), discriminates strongly among generator treatments of low hadronic energy. The general overprediction of the \psipvis{} peak position points to a systematic underestimation of the average nuclear missing energy; the Relativistic Mean Field approach (\texttt{NEUT-EDRMF}) is the clear exception. Second, the \psipvis{} distribution for $\Sigma T_p < 20$~MeV (Fig.~\ref{fig:first_sumtp_bin}) shows that most generators fail to describe the low-proton-momentum region. Only \texttt{NEUT-EDRMF}, the sole model deriving Pauli blocking self-consistently from quantum mechanics rather than through simple momentum cuts, reproduces the data there, indicating that first-principles treatments are essential for precise cross-section predictions at low proton kinetic energy. Third, tuning of the interaction model (\texttt{MnvTune4.4.1}) on lepton and hadron kinematics yields a reasonable prediction of the nuclear missing energy, despite a simplistic nuclear model that fails to describe other data~\cite{MINERvA:2019ope}. These results establish \psipvis{} as a tool to probe and reduce nuclear-model biases in the reconstruction of neutrino energy from visible final-state kinematics in quasielastic neutrino--nucleus interactions, of direct relevance to the measurement of neutrino oscillation parameters.

% Measurement of \psipvis{} allows to investigate the ability of a neutrino-interaction model to predict the nuclear missing energy in QE scattering and therefore provides an important tool to understand and reduce nuclear model-induced biases in the measurement of oscillation parameters. Analysis of the \psipvis{} peak position is found to discriminate strongly among model predictions, with most generator exhibiting too much absolute displacement (Fig.~\ref{fig:peaks_pt} and Fig.~\ref{fig:peaks_sumtp}) at low momentum transfer, with the exception of the Relativistic Mean Field approach. At the same time, tuning of the interaction model (\texttt{MnvTune4.4.1}) on observations of lepton and hadron kinematics yields reasonable prediction of the missing energy due to nuclear effects, despite using a simplistic nuclear model that fails to describe other data~\cite{MINERvA:2019ope}. 

% MnvTUne4.4.1 "despite not describing well other feetures, and begin very simole, this measurements show that the ref model ha sa good description of the partition ebtween visible and invisible energy." 

% acknowledgements

\FloatBarrier

\begin{acknowledgments}

This document was prepared by members of the MINERvA Collaboration using the resources of the Fermi National Accelerator Laboratory (Fermilab), a U.S. Department of Energy, Office of Science, Office of High Energy Physics HEP User Facility. Fermilab is managed by Fermi Forward Discovery Group, LLC, acting under Contract No. 89243024CSC000002.
These resources included support for the MINERvA construction project, and support
for construction also
was granted by the United States National Science Foundation under
Award No. PHY-0619727 and by the University of Rochester. Support for
participating scientists was provided by NSF and DOE (USA); by CAPES
and CNPq (Brazil); by CoNaCyT (Mexico); by ANID PIA / APOYO AFB180002, CONICYT PIA ACT1413, and Fondecyt 3170845 and 11130133 (Chile); 
%by CONCYTEC, DGI-PUCP, and IDI/IGI-UNI (Peru); 
by CONCYTEC (Consejo Nacional de Ciencia, Tecnolog\'ia e Innovaci\'on Tecnol\'ogica), DGI-PUCP (Direcci\'on de Gesti\'on de la Investigaci\'on  - Pontificia Universidad Cat\'olica del Peru), and VRI-UNI (Vice-Rectorate for Research of National University of Engineering) (Peru); NCN Opus Grant No. 2016/21/B/ST2/01092 (Poland); by Science and Technology Facilities Council (UK); by EU Horizon 2020 Marie Skłodowska-Curie Action; by a Cottrell Postdoctoral Fellowship from the Research Corporation for Scientific Advancement; by an Imperial College London President's PhD Scholarship.  We thank the MINOS Collaboration for use of its near detector data. Finally, we thank the staff of
Fermilab for support of the beam line, the detector, and computing infrastructure. This work was partially funded by the Swiss National Science Foundation, grants No. 200021\_204609 and IZSEZ0\_233057.

\end{acknowledgments}

\bibliography{biblio_2}% Produces the bibliography via BibTeX.

%%% End matter %%%%

\clearpage
\raggedbottom % supposd to prevent LaTeX from stretching vertical space to fill columns evenly.  But doesn't seem to work

\section{End Matter}

\subsection{Description of benchmarked models}
\label{app:models}
Measurements are compared to predictions of modern neutrino--nucleus interaction models, produced with different event generators (NEUT~\cite{NEUT}, GENIE~\cite{AndreopoulosGENIE2010}, and NuWro~\cite{NuWro}). The processes contributing to QE-Like neutrino--nucleus scattering at MINERvA energies are, other than genuine QE interaction: two-particle-two-holes (2p2h) processes, pion production from baryon resonances (RES), and deep inelastic scattering (DIS) where any produced hadrons that are not nucleons have been absorbed in the nucleus. Typically, neutrino event generators simulate different processes independently, at times even using profoundly different nuclear models for different processes. Table~\ref{tab:mc_models} summarizes the theoretical model underlying each of the relevant processes for all models benchmarked in this analysis. The first column states the model name as used in the Figs.~\ref{fig:MnvTune_Pt}--\ref{fig:first_sumtp_bin}, always including the generator name (with the exception of the reference model, \texttt{MnvTune4.4.1}, using GENIE v2.12.2). All models use the BBBA05 parameterization of the nucleon vector form factor~\cite{BRADFORD2006127}, and a dipole parametrization of the axial form factor, with the exception of the reference model which uses a $z$-expansion parameterization of $\nu D$ scattering data~\cite{Zexpansion}.

The model \texttt{GENIE-v3} is based on the GENIE configuration denoted as G18\_10b\_02\_11a~\cite{genie_tunes}, using GENIE version 3.0.6. The predictions labelled \texttt{GENIE-v3} in this paper use the QE and 2p2h \texttt{GENIE-v3} implementation, while the resonant and DIS are kept as the reference model \texttt{MnvTune4.4.1}, due to a mismodelling of resonance decay in GENIE v3.0.6.generating unphysical abundance of events at low $\Sigma T_p$.

Several modifications (``tunes'') are applied to the reference model \texttt{MnvTune4.4.1} with respect to the description in Table~\ref{tab:mc_models}. Long-range nucleon correlations are included through the random-phase approximation (RPA) calculations of the Valencia group~\cite{PhysRevC.70.055503,gran2017modeluncertaintiesvalenciarpa}, leading to a suppression of the quasielastic cross-section at low $Q^2$. The multinucleon knockout contribution (mainly 2p2h) is increased in order to account for Data/MC mismatch observed in low energy (LE) beam MINERvA data~\cite{PhysRevLett.116.071802}. Additionally, the single pion production channel is tuned to a recent re-analysis of bubble-chamber neutrino-deuteron interactions~\cite{rodriguesConstrainingGENIEModel2016}, resulting in a significant reduction of the non-resonant pion production. These {\it ad hoc} modifications to the model represent an attempt to reproduce as accurately as possible previous MINERvA measurements not directly related to \psipvis{}~\cite{collaborationSimultaneousMeasurementProton2022}.
%, and are detailed in~\cite{collaborationSimultaneousMeasurementProton2022}.

\begin{table*}[!]
\centering
\caption{Summary of the interaction models used in the analysis. For each model, the modelling choices for the main interaction channels and final-state interactions are reported.}
\label{tab:mc_models}
\adjustbox{max width=\textwidth}{%
\begin{tabular}{llllll}
\toprule
\textbf{Model} & \textbf{Nuclear state} & \textbf{2p2h} & \textbf{Resonant} & \textbf{DIS / Multi-pion} & \textbf{Intranuclear FSI} \\
\midrule
\texttt{MnvTune4.4.1} 
  & Global Fermi gas~\cite{SMITH1972605} & Valencia~\cite{VALENCIA} & Rein-Sehgal~\cite{ReinSehgal} & AGKY hadronization~\cite{Yang_2009} & \texttt{INTRANUKE-hA} empirical  \\
  & with Bodek-Ritchie tail~\cite{BR_tail} &  & $M_A^{RES}=1.12$~GeV/$c^2$ & &  single-step cascade \\
\midrule
\texttt{NEUT-SF} 
  & Rome spectral function~\cite{BENHAR1994493} & Valencia & Rein-Sehgal & $W>2$~GeV: PYTHIA~\cite{PYTHIA5} + Bodek-Yang~\cite{BY_DIS} & Semi-classical cascade \\
  &  & & $M_A^{RES}=1.21$~GeV/$c^2$ & $W<2$~GeV: tuned to bubble chamber data~\cite{NEUT} & (0.2~fm steps) \\
\midrule
\texttt{NEUT-EDRMF} 
  & Energy-dependent & Valencia & Rein-Sehgal & $W>2$~GeV:PYTHIA + Bodek-Yang & Semi-classical cascade \\
  & relativistic mean field~\cite{jake_EDRMF} & & $M_A^{RES}=1.21$~GeV/$c^2$ & $W<2$~GeV: tuned to bubble chamber data & (0.2~fm steps)\\
\midrule
\texttt{NuWro-LFG} 
  & Local Fermi gas~\cite{bourguilleInclusiveExclusiveNeutrinonucleus2021} & Valencia & Ghent hybrid & $W>2$~GeV: PYTHIA + Bodek-Yang & Semi-classical cascade \\
  &  & & model~\cite{yanGhentHybridModel2024} &  $W<2$~GeV: NuWro hadronization model~\cite{NuWro}& (0.2~fm steps) \\
\midrule
\texttt{GENIE-v3} 
  & Local Fermi gas~\cite{valverdeQuasielasticNeutrinonucleusReactions2006} & Valencia & Berger-Sehgal~\cite{berger_sehgal} & AGKY hadronization& \texttt{INTRANUKE-hN} empirical \\
  &  & & & & cascade (0.2~fm steps) \\
\bottomrule
\end{tabular}%
}
\end{table*}

\subsection{Forward-folding analysis and systematic uncertainties}
\label{app:howto}
The recent release of MINERvA data~\cite{opendata} enables comparison with a wide range of model predictions. In this appendix, we describe the forward-folding method used to obtain such comparisons at reconstructed level without regenerating full detector-simulation samples, as presented in this Letter.

In this analysis, measured binned distributions, necessarily convoluted with detector effects, are compared with interaction model predictions on which detector effects have been simulated. 

It is common practice in neutrino physics experiments to build a complete detector model to simulate the effect of the detector response on kinematic distributions predicted by a reference neutrino--nucleus interaction model. Producing simulated detector response for many different models is often computationally expensive, and experiments most often fully simulate only one interaction model. In this analysis, the tool described in Ref.~\cite{Ziggy} is used to obtain reconstruction-level distributions and uncertainties for any interaction model and using MINERvA detector simulation. It employs boosted decision trees (BDTs)~\cite{BDT_general} to derive event-by-event weights that transform one simulated neutrino--nucleus interaction dataset (the source) into another (the target), matched in terms of a chosen set of analysis variables. In this analysis, a BDT is trained to reweight the MINERvA reference model to four different target samples in terms of the three analysis variables $p_{\text{T}\mu}$, $\Sigma T_p$, and \psipvis{}.  Importantly, since the weights are derived jointly from all three analysis variables, the reweighted sample also captures the correlations among them.
%, and not merely their individual one-dimensional projections. 
Different processes contributing to the signal sample (QE, 2p2h and ``Others'', including both DIS and pion production) are reweighted individually and independently of each other, to retain the relative differences between irreducible backgrounds and signal.
The BDT is trained using a source and a target model sample with sufficient statistics, and it produces a reweighter object that computes the event weight as a function of the true event kinematics for each process: \( w(p_{\text{T}\mu},\Sigma T_p,\psi_{\rm vis}^\prime;{\rm Process})\). 
The resulting event weights are applied at event selection time, yielding reconstruction-level distributions for the target interaction model.
The tool for creating weights is open source and public~\cite{git_repo_bdtreweight}.
 
Systematic uncertainties are estimated using the multiuniverse method, as in other MINERvA analyses~\cite{minerva_multiuniverse}. In this approach, the baseline MC prediction is varied according to a set of plausible model variations, and each varied model is used to fill a separate ``universe'', i.e., a histogram corresponding to that particular variation. This procedure is repeated independently for each source of uncertainty yielding a collection of universes for each source from which the corresponding uncertainty is estimated. 

Within the forward-folding method, uncertainties have to be estimated in the reconstructed space for both Data and MC predictions. We are interested specifically in the uncertainty induced by the detector response, including detector response variations caused by variations of the interaction model. For each analysis bin in the three-dimensional space spanned by the analysis variables $p_{\text{T}\mu}$, $\Sigma T_p$, and \psipvis{}, we compute from the MINERvA reference model the efficiency $E$, and the migration matrix $U$, which maps event migrations from the true bins to the reconstructed bins. For a given uncertainty source $m$ and universe $n$, the corresponding efficiency and migration matrix are denoted $E_m^n$ and $U_m^n$, respectively. The propagated reconstructed-level universe is then obtained as
\begin{equation}
R_m^n = U_m^n \left( E_m^n \odot T \right),
\label{eq:multiuniverse_forward_fold}
\end{equation}
where $\odot$ denotes an element-wise (Hadamard) product and the multiplication by $U_m^n$ is the standard matrix multiplication. Here $T$ is the truth-level central-value histogram, common to all universes of all uncertainty sources, so that only the model-induced variations of $E$ and $U$ propagate into the reconstructed-level uncertainty. This method allows to propagate to the reconstructed space the systematic uncertainties caused by variations of the interaction model, by selecting only the effect that the interaction model variation has on the detector response, correctly describing the measurement's systematic uncertainties. 

\paragraph{\psipvis{} peak estimation}

The \psipvis{} peak position is measured via kernel density estimation on the reconstruction-level histograms. The peak is identified by the maximum of the smooth function obtained by summing $N_{\text{bins}}$ Gaussian distributions with mean at the bin center and $\sigma$ proportional to the bin width, each weighted by the bin content. We adopt $\sigma=0.6\times\text{bin width}$, chosen to balance the bias–variance trade-off: small enough to retain the relevant features of the distribution, yet large enough to avoid overfitting the bin-to-bin statistical noise. Statistical uncertainties are obtained by repeating the peak estimation on 20 replicas of the histogram, in which each bin content is fluctuated according to a Poisson distribution, and taking the standard deviation of the resulting peak positions. Systematic uncertainties are estimated by recomputing the peak position in each of the systematic universes, and then extracting the uncertainty as the standard deviation of the obtained peak positions. 

\end{document}

\preprint{FERMILAB-PUB-26-0707-PPD}

\title{Supplemental Materials}

% \author{Ann Author}

% \collaboration{MINER$\nu$A Collaboration}%\noaffiliation

\maketitle

\section{Presentation of the \psipvis{} distributions as a function of the final state kinematics for all models under study}
Fig.~\ref{fig:allmodels_Pt} shows the measured \psipvis{} distribution on the signal as a function of the muon transverse momentum $p_{\text{T}\mu}$, while Fig.~\ref{fig:allmodels_SumTp} shows it as a function of the hadronic recoil $\Sigma T_p$, for all the models benchmarked in the manuscript. The \psipvis{} peak position shown and discussed in the manuscript is estimated from these histograms. Error bars are reported on the data histogram and include both statistical and systematic errors affecting the signal and background samples, summed in quadrature. 

We also report the data-simulation comparison in reconstructed space with interaction-mode breakdown for each of the benchmarked models, in the format of Figures 1 and 2 of the paper, to show how the event rates predictions are subdivided between true QE events, Resonant pion production, 2p2h and DIS. The reconstructed \psipvis{} distribution breakdown predictions are shown in Figs.~\ref{fig:pt_neutsf} and~\ref{fig:sumtp_neutsf} for \texttt{NEUT-SF}, in Figs.~\ref{fig:pt_neutedrmf} and~\ref{fig:sumtp_neutedrmf} for \texttt{NEUT-EDRMF}, in Figs.~\ref{fig:pt_NuWro-LFG} and~\ref{fig:sumtp_NuWro-LFG} for \texttt{NuWro-LFG}, and in Figs.~\ref{fig:pt_GENIEv3} and~\ref{fig:sumtp_GENIEv3} for \texttt{GENIEv3}.

\begin{figure*}[p]
    \centering
    \includegraphics[width=0.8\linewidth]{SM_Figs/AllModels_poly/AllSlices_Pt_5x3_poly.png}
    \caption{Reconstructed \psipvis{} distribution in bins of the muon transverse momentum $p_{T\mu}$.}
    \label{fig:allmodels_Pt}
\end{figure*}

\begin{figure*}[p]
    \centering
    \includegraphics[width=0.8\linewidth]{SM_Figs/AllModels_poly/AllSlices_SumTp_4x3_poly.png}
    \caption{Reconstructed \psipvis{} distribution in bins of the hadronic recoil $\Sigma T_p$.}
    \label{fig:allmodels_SumTp}
\end{figure*}

\begin{figure*}[tp]
    \centering
    \includegraphics[width=0.9\linewidth]{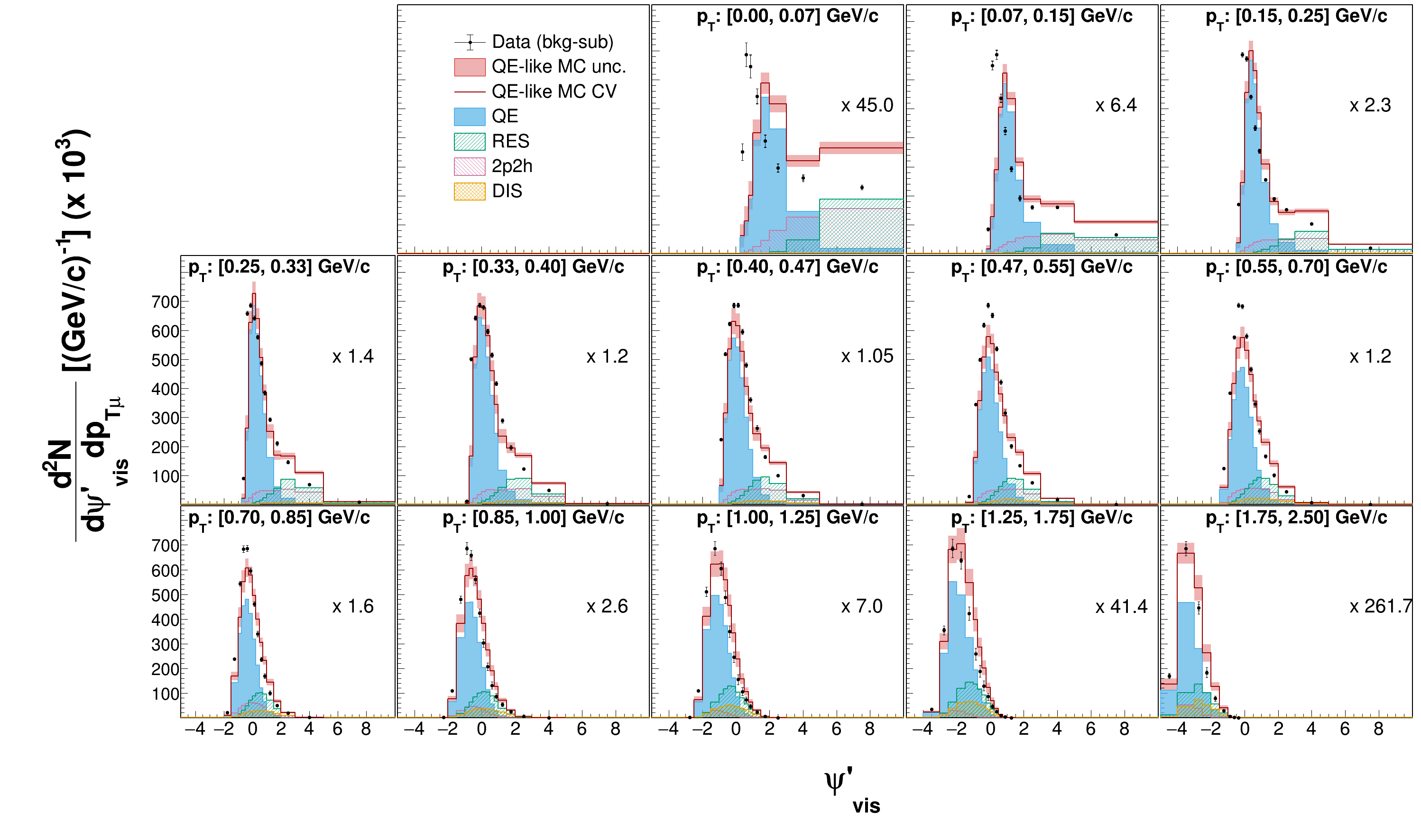}
    \caption{Reconstructed \psipvis{} distribution as a function of $p_{T\mu}$ and interaction-mode breakdown compared to data for \texttt{NEUT-SF}.}
    \label{fig:pt_neutsf}
\end{figure*}
\begin{figure*}[tp]
    \centering
    \includegraphics[width=0.9\linewidth]{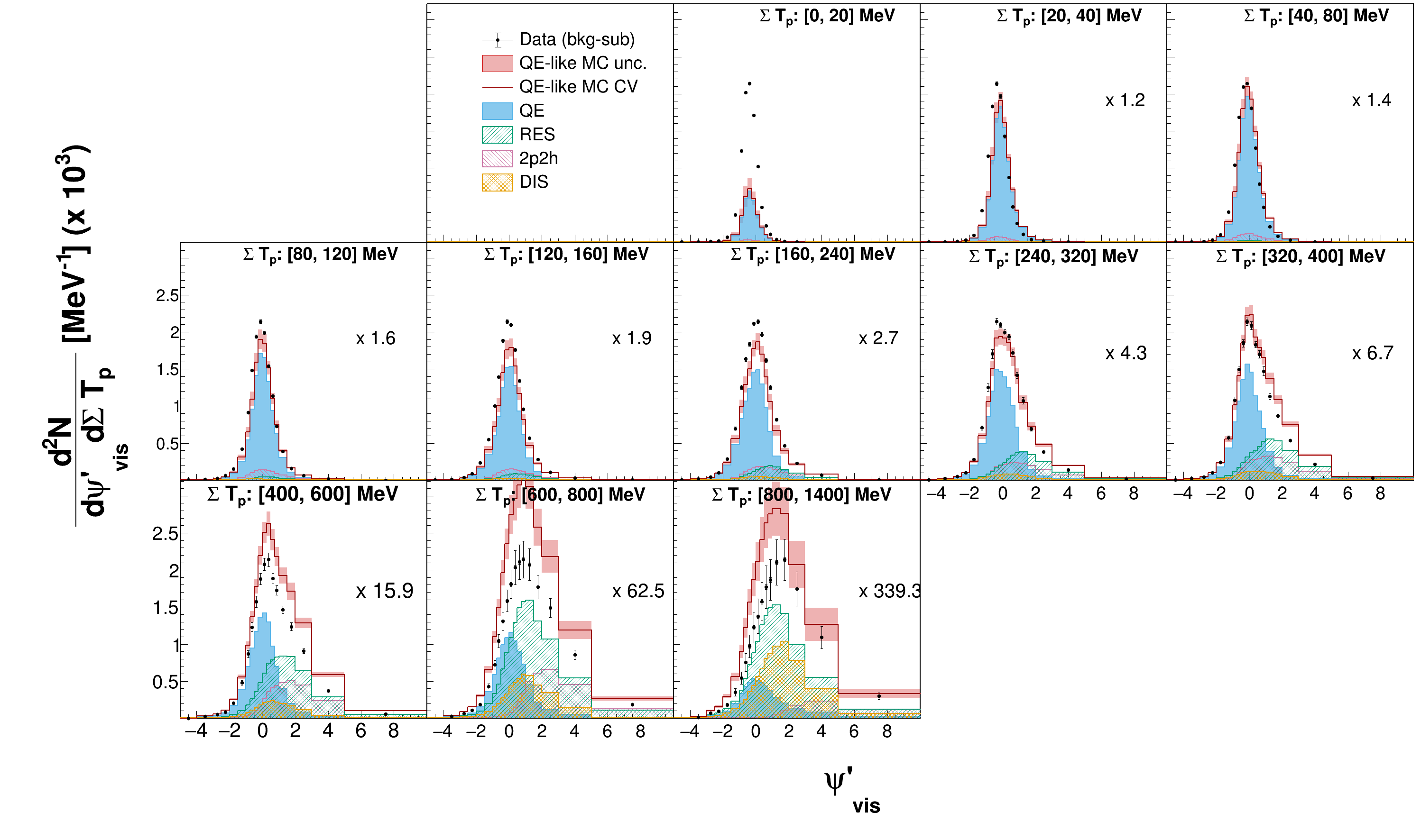}
    \caption{Reconstructed \psipvis{} distribution as a function of $\Sigma T_p$ and interaction-mode breakdown compared to data for \texttt{NEUT-SF}.}
    \label{fig:sumtp_neutsf}
\end{figure*}

\begin{figure*}[tp]
    \centering
    \includegraphics[width=0.9\linewidth]{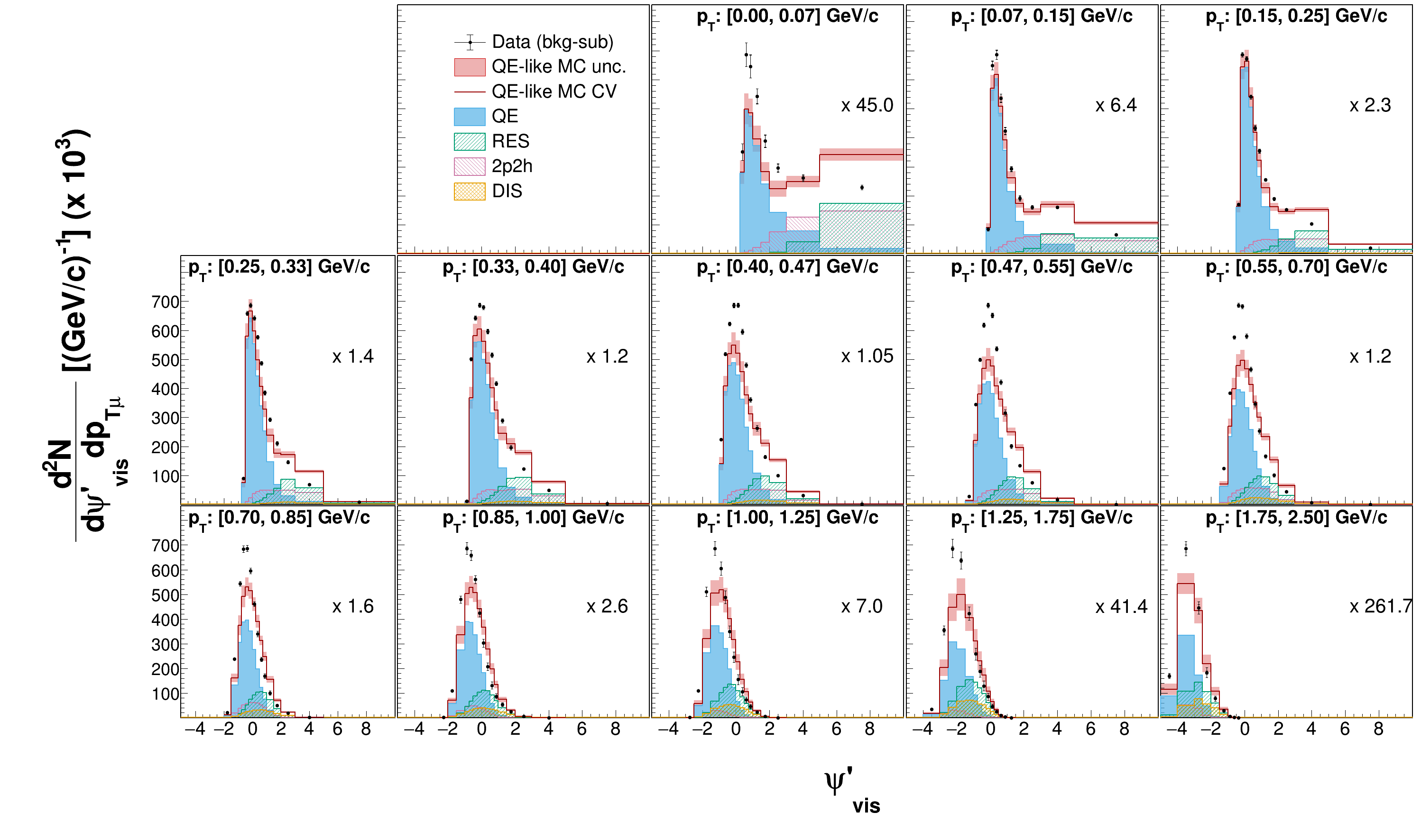}
    \caption{Reconstructed \psipvis{} distribution as a function of $p_{T\mu}$ and interaction-mode breakdown compared to data for \texttt{NEUT-EDRMF}.}
    \label{fig:pt_neutedrmf}
\end{figure*}
\begin{figure*}[tp]
    \centering
    \includegraphics[width=0.9\linewidth]{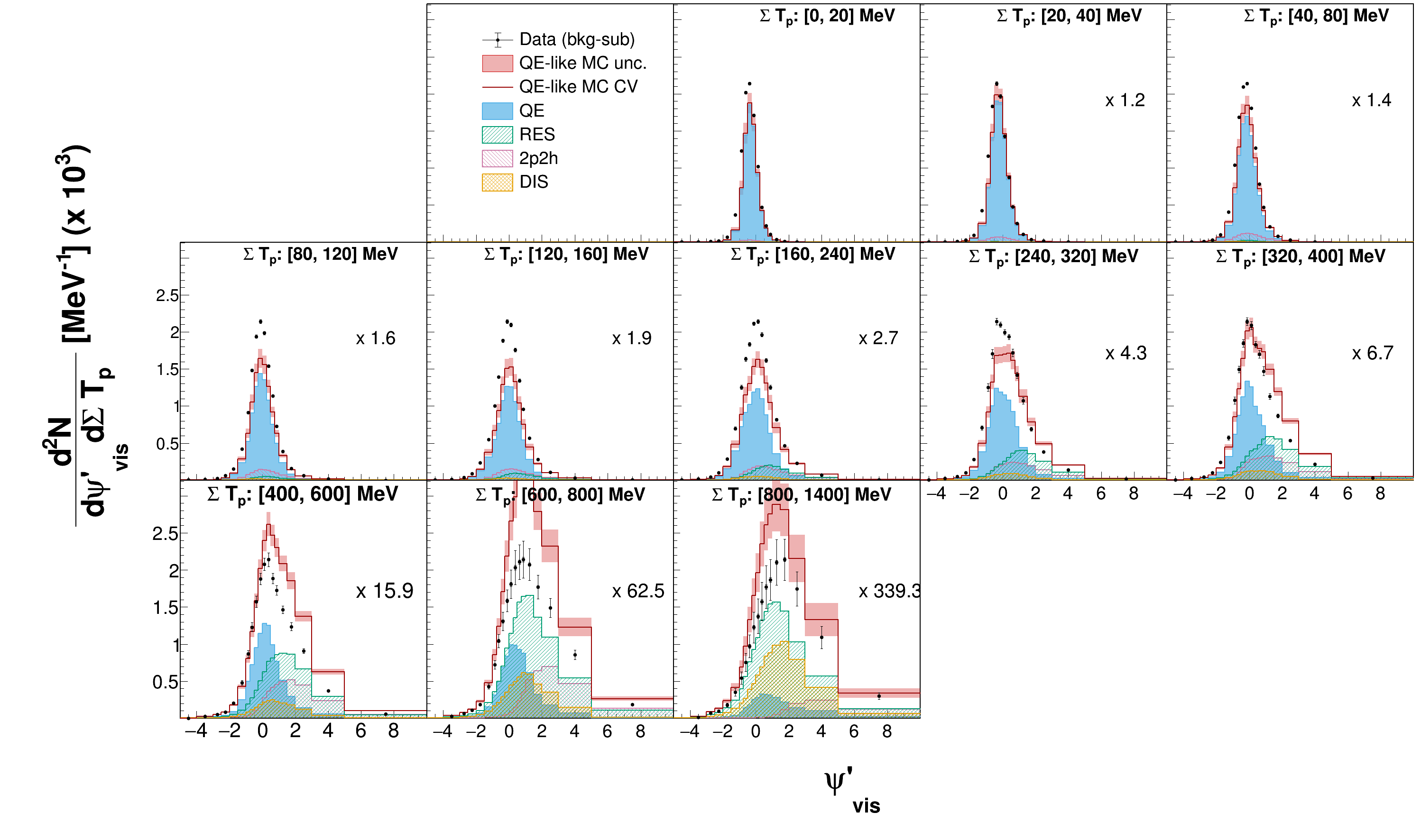}
    \caption{Reconstructed \psipvis{} distribution as a function of $\Sigma T_p$ and interaction-mode breakdown compared to data for \texttt{NEUT-EDRMF}.}
    \label{fig:sumtp_neutedrmf}
\end{figure*}

\begin{figure*}[tp]
    \centering
    \includegraphics[width=0.9\linewidth]{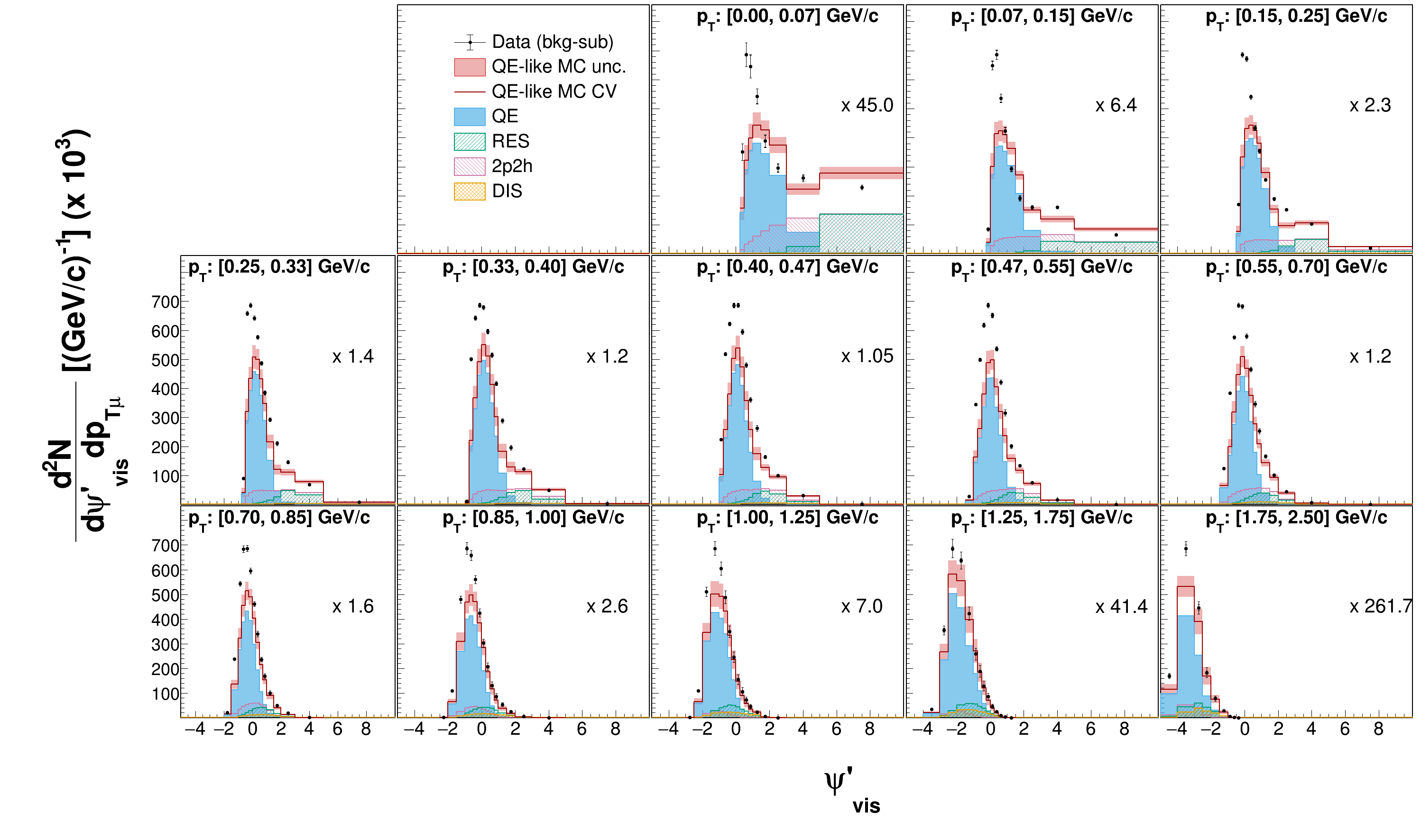}
    \caption{Reconstructed \psipvis{} distribution as a function of $p_{T\mu}$ and interaction-mode breakdown compared to data for \texttt{NuWro-LFG}.}
    \label{fig:pt_NuWro-LFG}
\end{figure*}
\begin{figure*}[tp]
    \centering
    \includegraphics[width=0.9\linewidth]{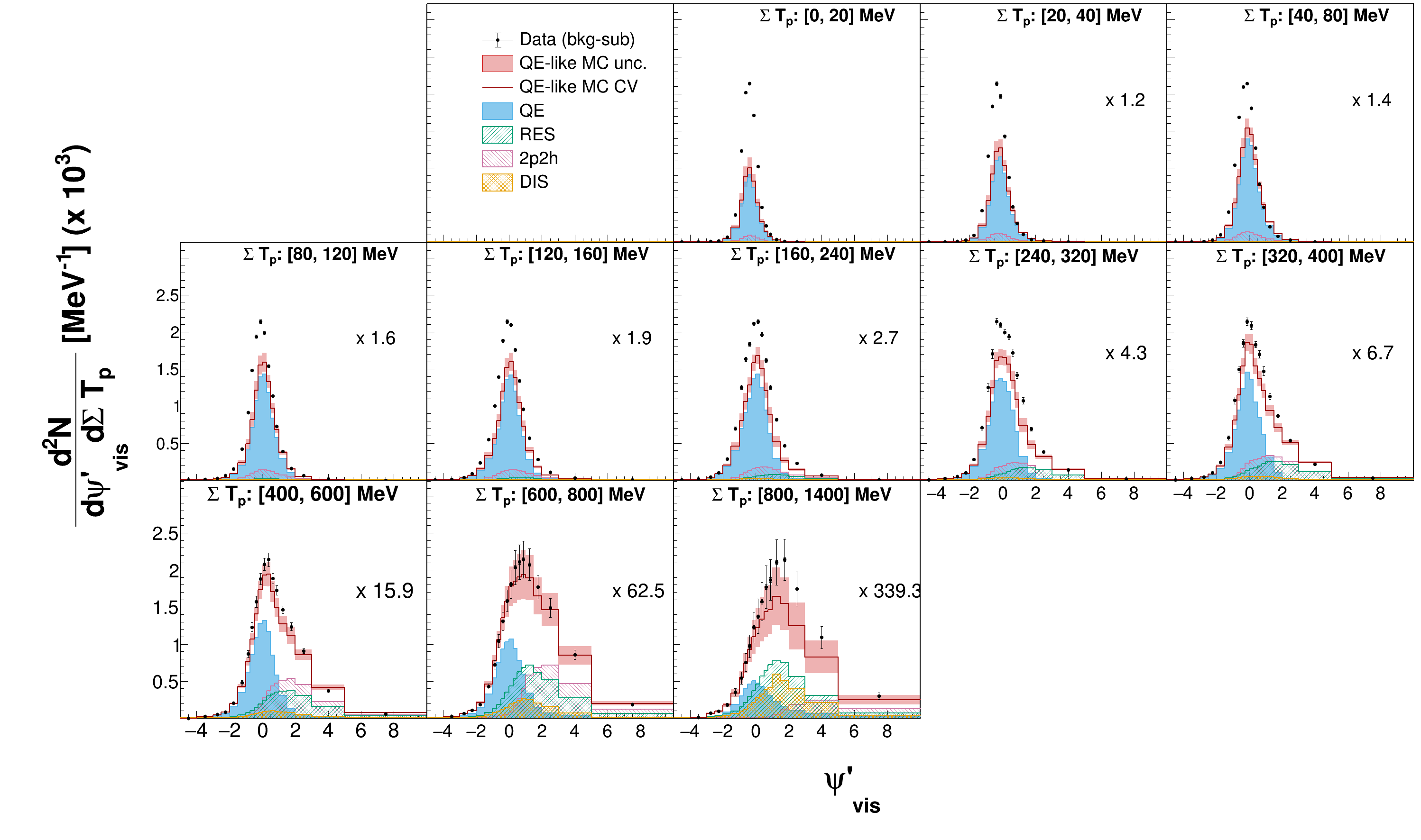}
    \caption{Reconstructed \psipvis{} distribution as a function of $\Sigma T_p$ and interaction-mode breakdown compared to data for \texttt{NuWro-LFG}.}
    \label{fig:sumtp_NuWro-LFG}
\end{figure*}

\begin{figure*}[tp]
    \centering
    \includegraphics[width=0.9\linewidth]{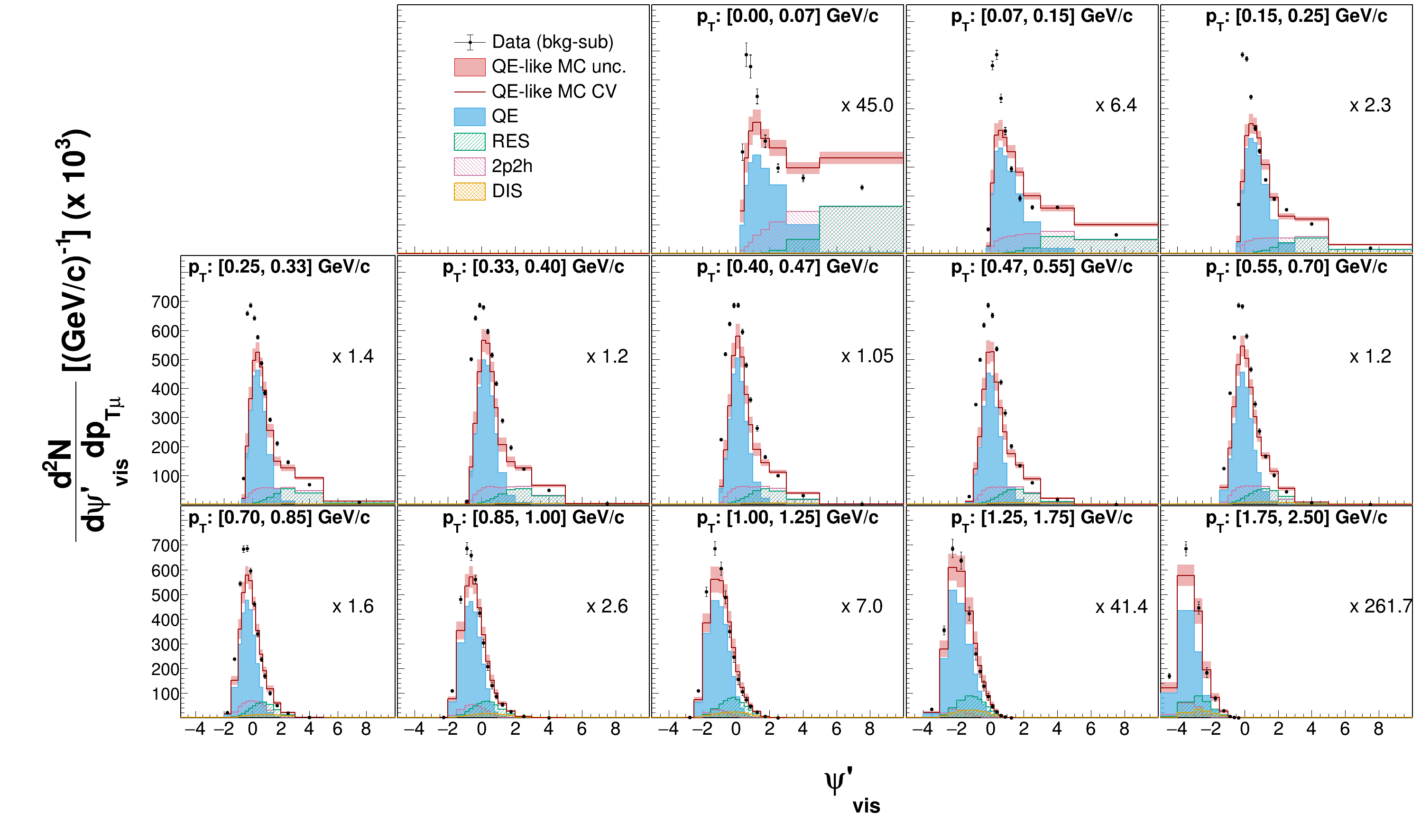}
    \caption{Reconstructed \psipvis{} distribution as a function of $p_{T\mu}$ and interaction-mode breakdown compared to data for \texttt{GENIEv3}.}
    \label{fig:pt_GENIEv3}
\end{figure*}
\begin{figure*}[tp]
    \centering
    \includegraphics[width=0.9\linewidth]{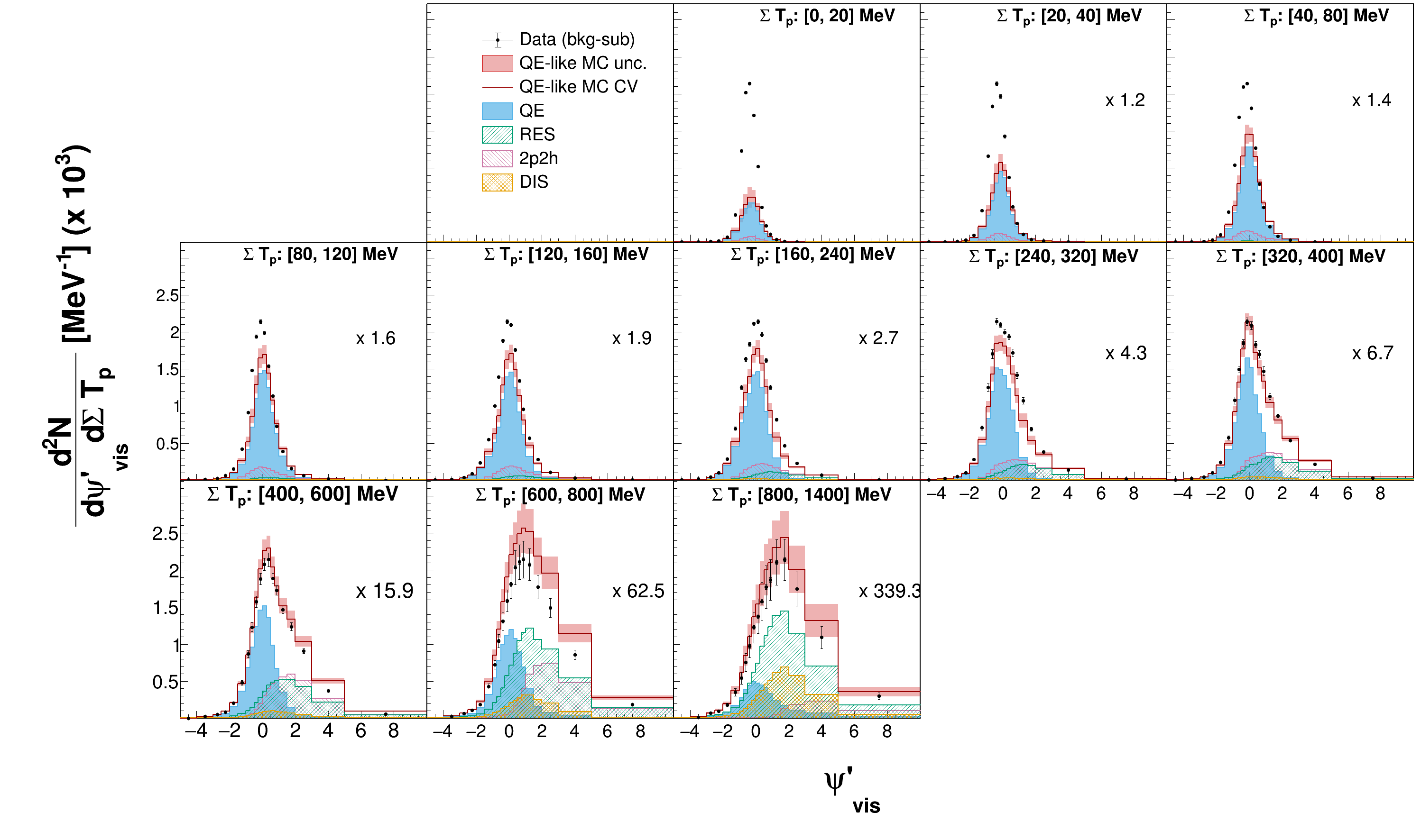}
    \caption{Reconstructed \psipvis{} distribution as a function of $\Sigma T_p$ and interaction-mode breakdown compared to data for \texttt{GENIEv3}.}
    \label{fig:sumtp_GENIEv3}
\end{figure*}

\clearpage

\section{Validation of the BDT reweighting method to reproduce model predictions}

Different model predictions are obtained by means of a reweighing method based on Boosted Decision Trees, as described in the End Matter. Event-by-event weights, derived by a BDT trained to reweight the source model's truth-level kinematics into the target model's, are applied to the already-simulated, reconstructed source-sample events, yielding reconstruction-level distributions for the target model without rerunning the detector simulation. Here we report the agreement between the BDT-reweighted source and the target distributions in the analysis bins at truth-level, to demonstrate validity of the reweighting method. Weights are computed independently for the three dominant contributions to CCQE-like sample: QE, 2p2h and pion/hadron production. We show the resulting cumulative CCQE-like distributions. 

\begin{figure*}
    \centering
    \includegraphics[width=1\linewidth]{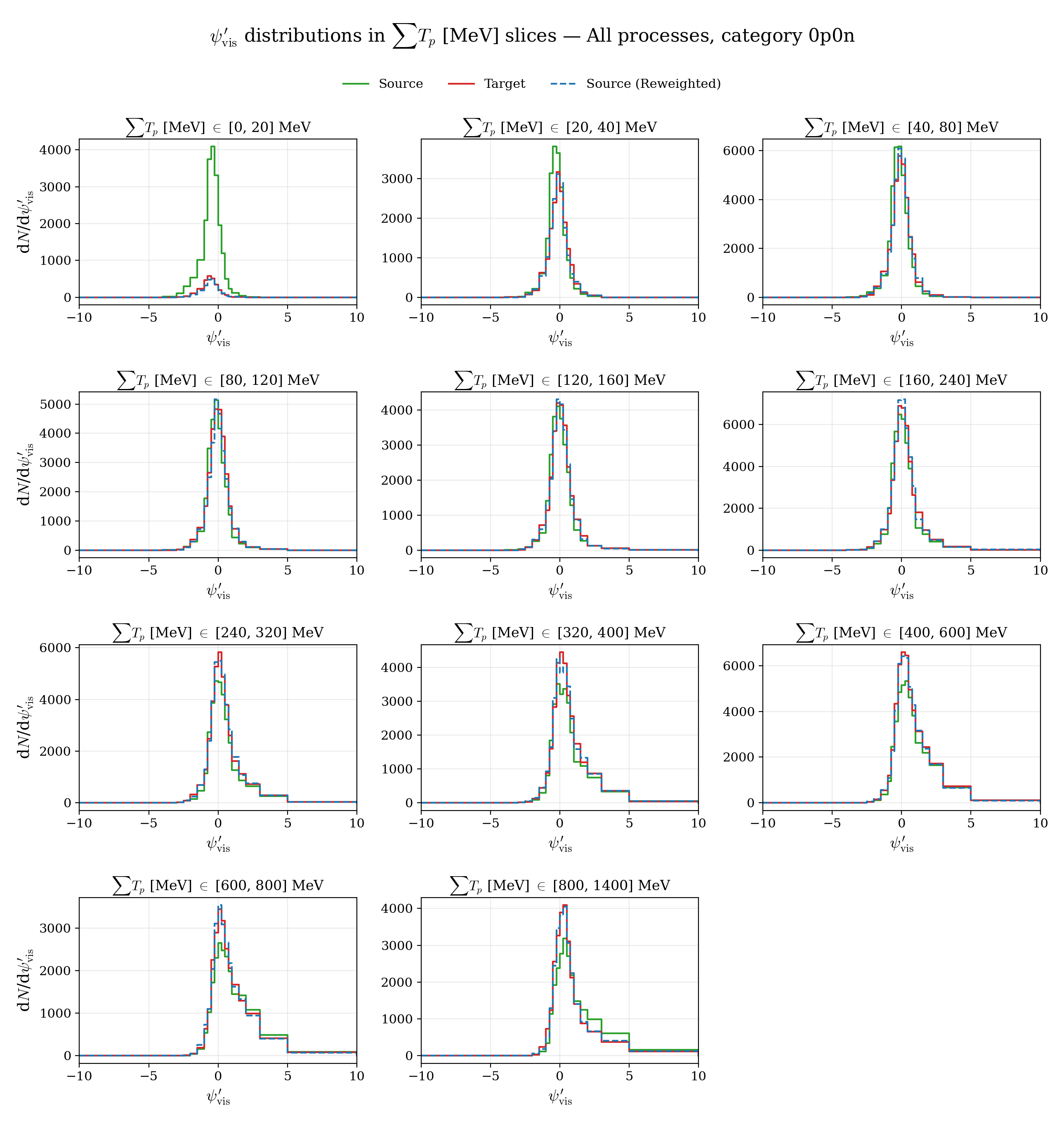}
    \caption{\texttt{NEUT-SF}, in bins of $\Sigma T_p$.}
    \label{fig:bdt_valid_neutsf_sumtp}
\end{figure*}

\begin{figure*}
    \centering
    \includegraphics[width=1\linewidth]{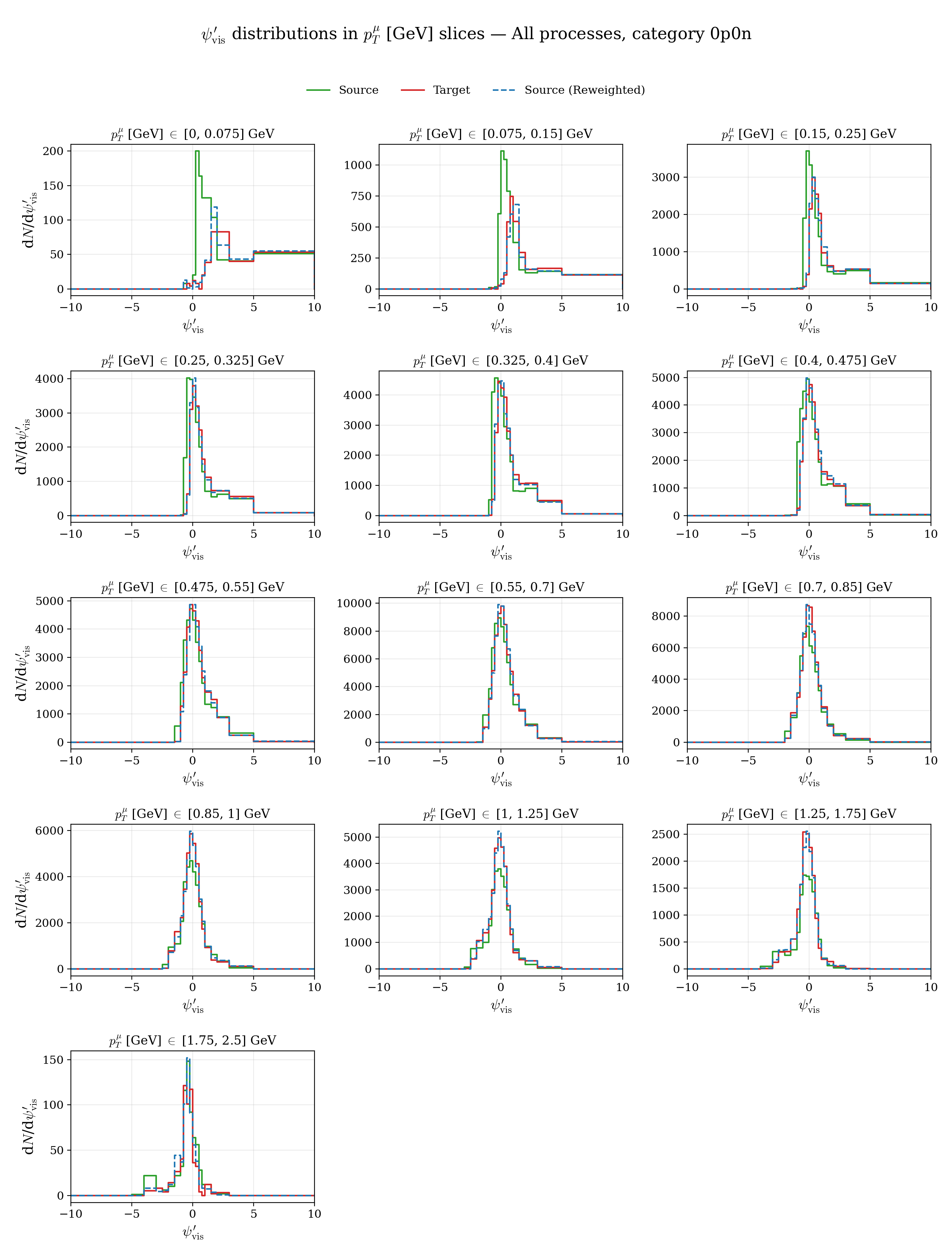}
    \caption{\texttt{NEUT-SF}, in bins of $p_{T\mu}$.}
    \label{fig:bdt_valid_neutsf_pt}
\end{figure*}

\begin{figure*}
    \centering
    \includegraphics[width=1\linewidth]{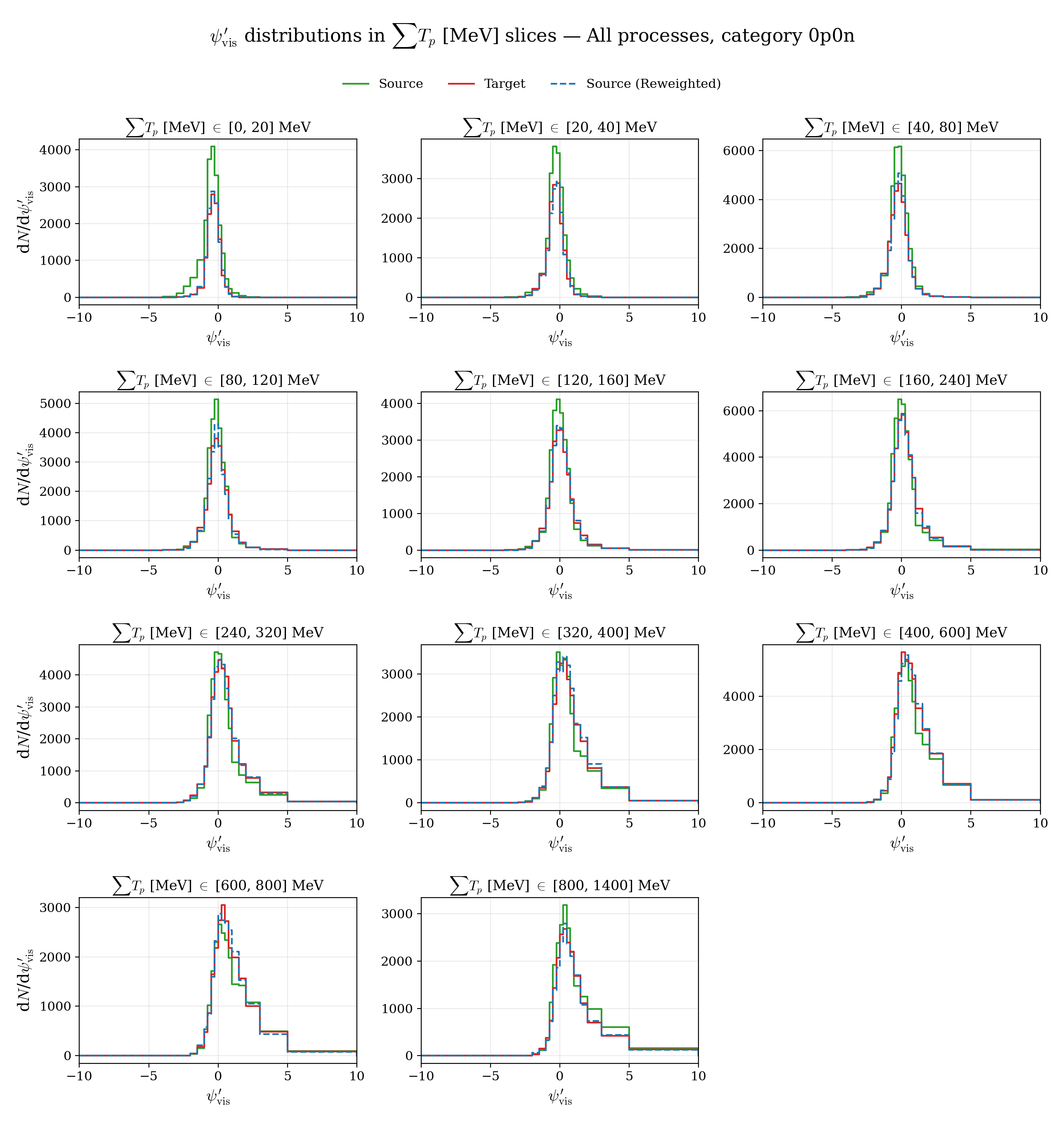}
    \caption{\texttt{NEUT-EDRMF}, in bins of $\Sigma T_p$.}
    \label{fig:bdt_valid_neutedrmf_sumtp}
\end{figure*}

\begin{figure*}
    \centering
    \includegraphics[width=1\linewidth]{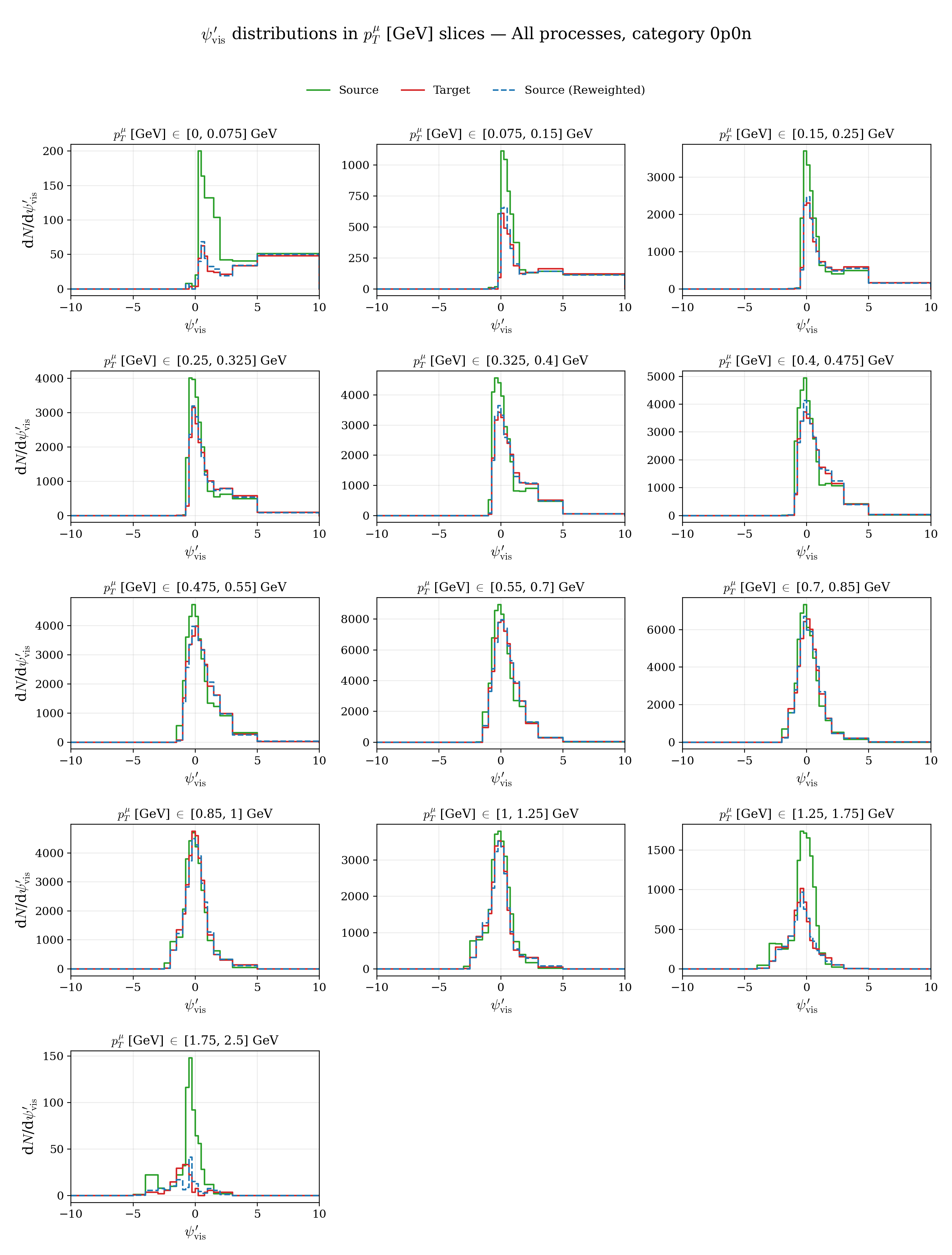}
    \caption{\texttt{NEUT-EDRMF}, in bins of $p_{T\mu}$.}
    \label{fig:bdt_valid_neutedrmf_pt}
\end{figure*}

\begin{figure*}
    \centering
    \includegraphics[width=1\linewidth]{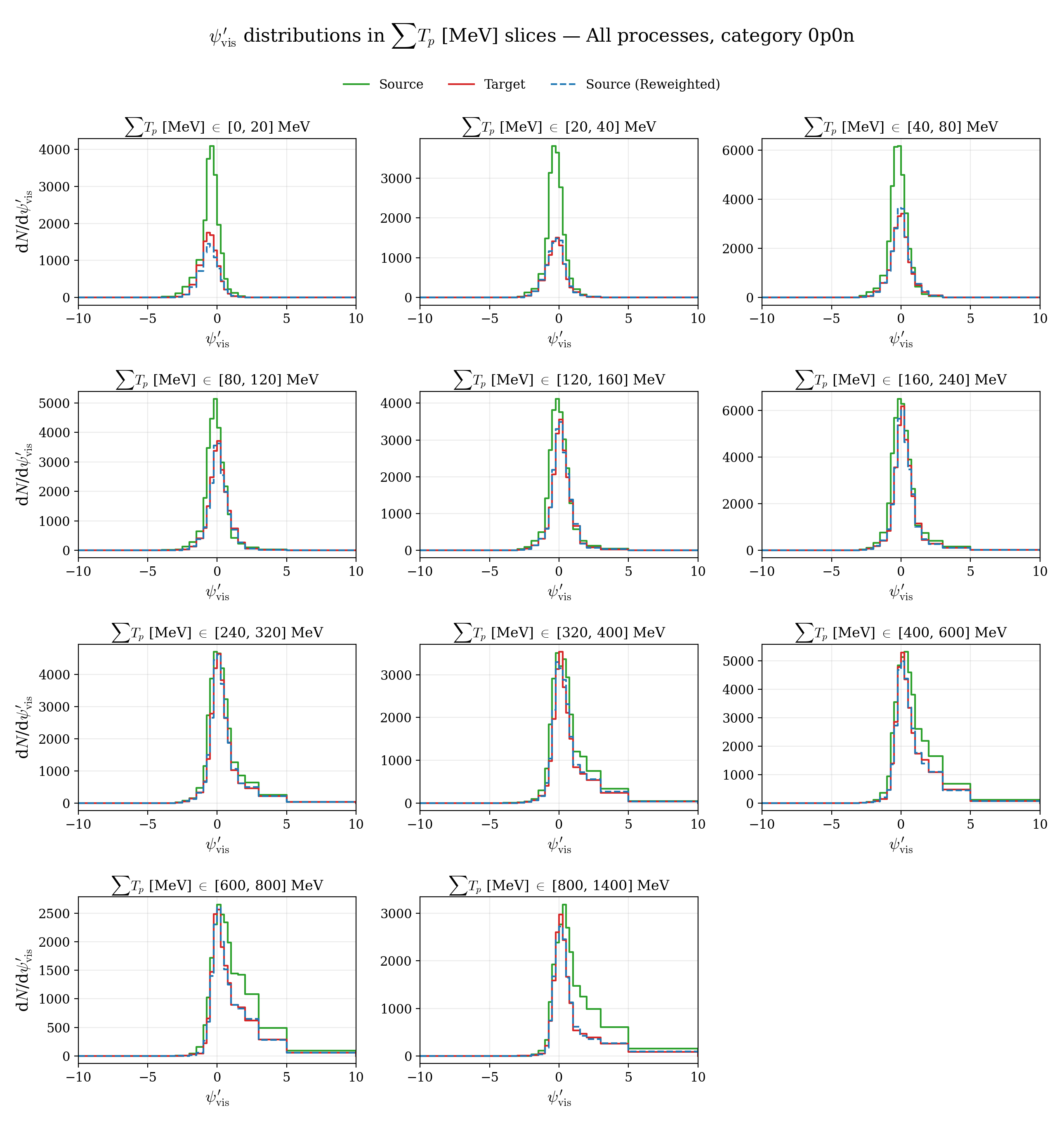}
    \caption{\texttt{NuWro-LFG}, in bins of $\Sigma T_p$.}
    \label{fig:bdt_valid_nuwro_sumtp}
\end{figure*}

\begin{figure*}
    \centering
    \includegraphics[width=1\linewidth]{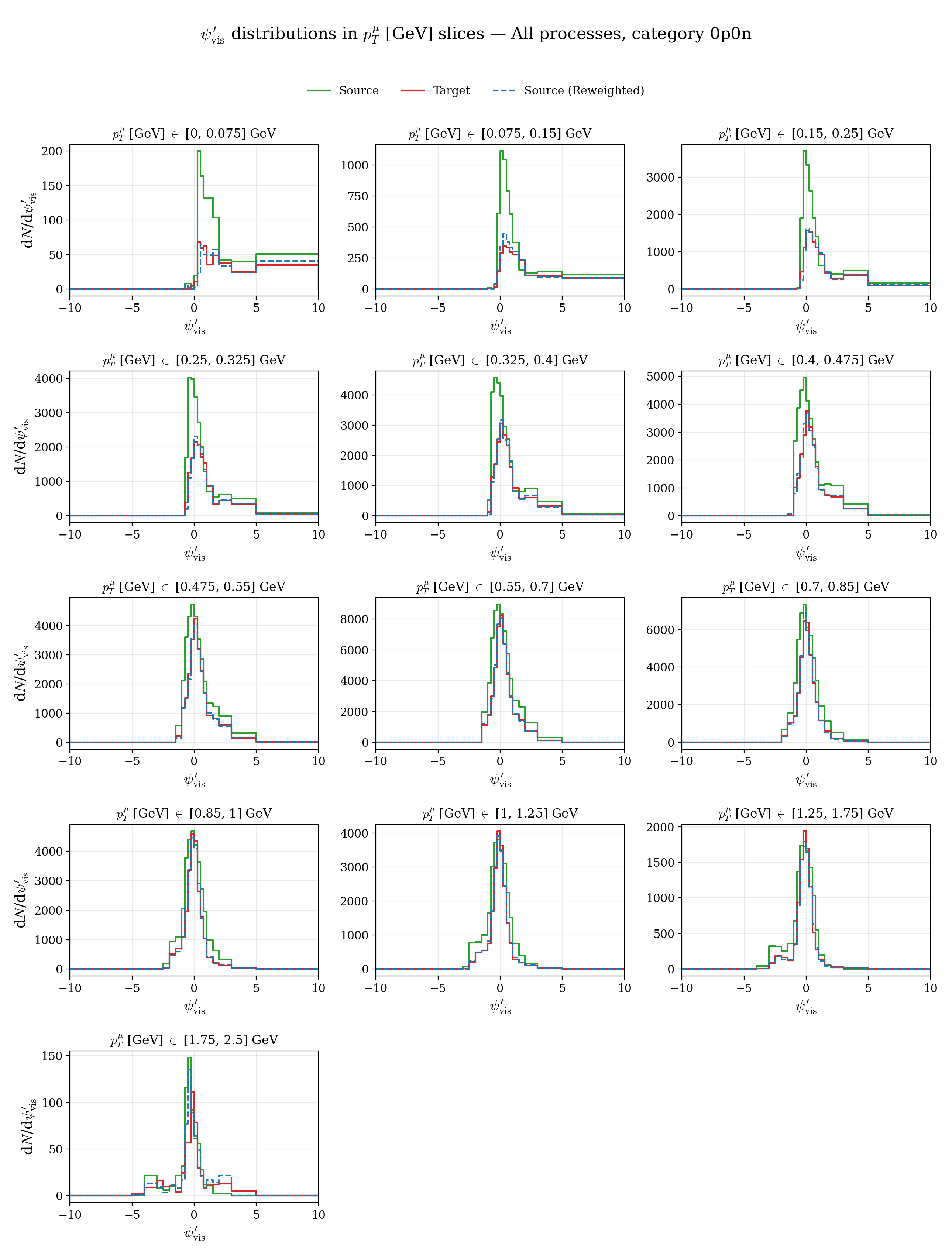}
    \caption{\texttt{NuWro-lFG}, in bins of $p_{T\mu}$.}
    \label{fig:bdt_valid_nuwro_pt}
\end{figure*}

\begin{figure*}
    \centering
    \includegraphics[width=1\linewidth]{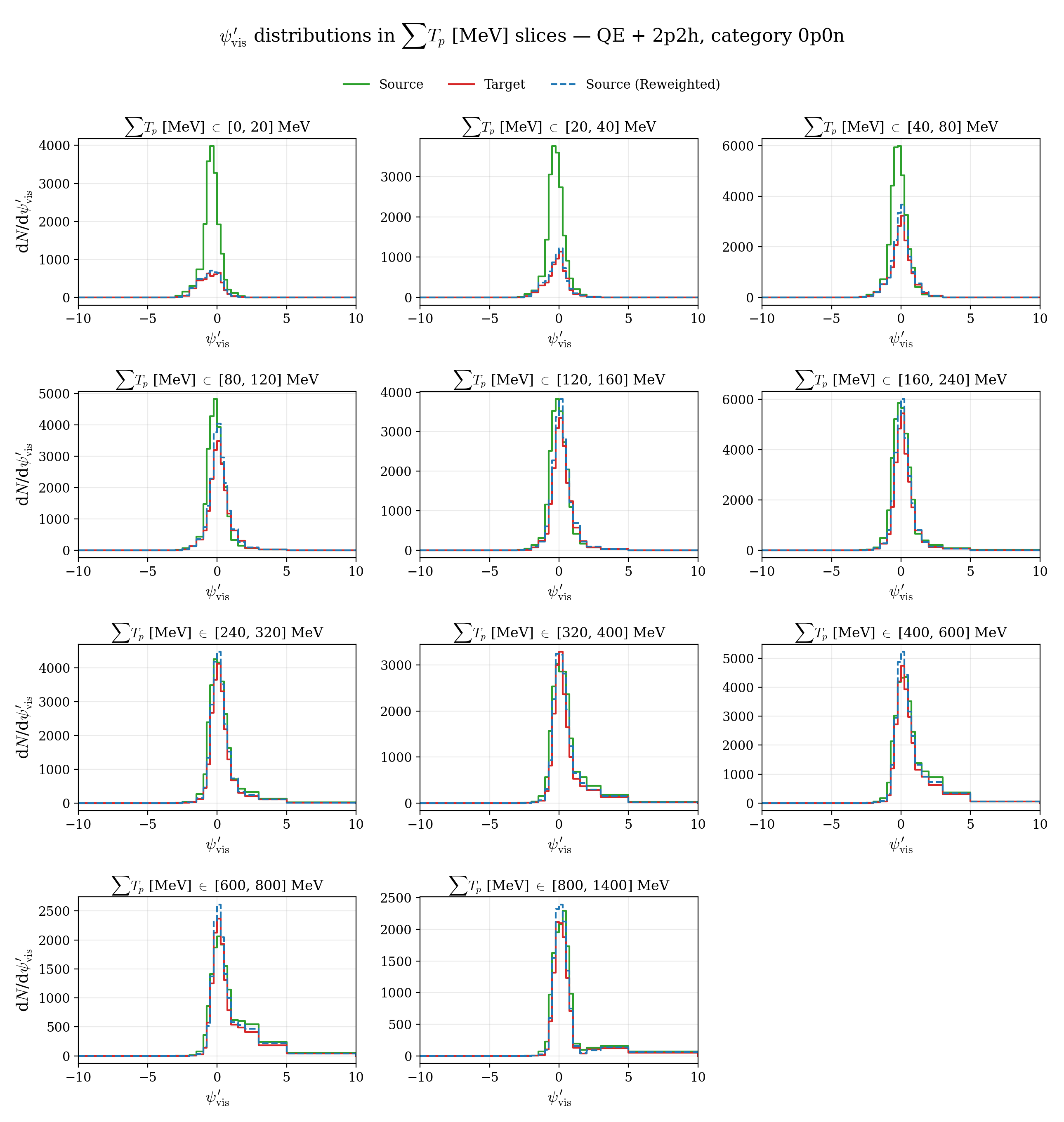}
    \caption{\texttt{GENIEv3}, in bins of $\Sigma T_p$. The resonant and DIS models are not reweighted due to a bug in the GENIE event generator.}
    \label{fig:bdt_valid_GENIEv3_sumtp}
\end{figure*}

\begin{figure*}
    \centering
    \includegraphics[width=1\linewidth]{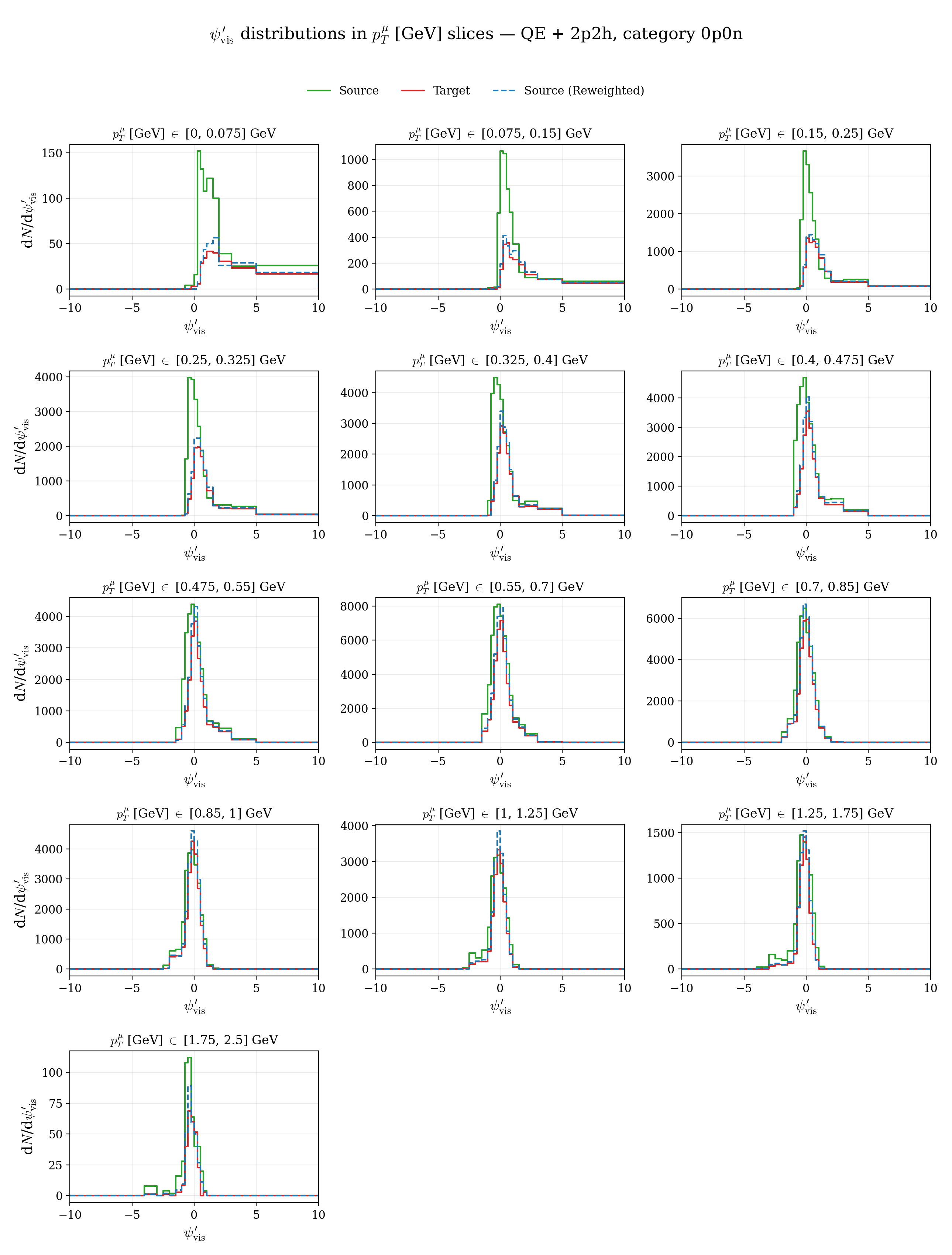}
    \caption{\texttt{GENIEv3}, in bins of $p_{T\mu}$. The resonant and DIS models are not reweighted due to a bug in the GENIE event generator. }
    \label{fig:bdt_valid_GENIEv3_pt}
\end{figure*}

Fig.~\ref{fig:bdt_valid_neutedrmf_pt} and Fig.~\ref{fig:bdt_valid_neutedrmf_sumtp} show the comparison between source (unweighted), source (reweighted) and target distributions for the \texttt{NEUT-EDRMF} model, respectively in bins of $p_{T\mu}$ and $\Sigma T_p$. Analogous plots are shown in Fig.~\ref{fig:bdt_valid_neutsf_pt} and Fig.~\ref{fig:bdt_valid_neutsf_sumtp} for the \texttt{NEUT-SF} model, Fig.~\ref{fig:bdt_valid_nuwro_pt} and Fig.~\ref{fig:bdt_valid_nuwro_sumtp} for the \texttt{NuWro-LFG} model, and Fig.~\ref{fig:bdt_valid_GENIEv3_pt} and Fig.~\ref{fig:bdt_valid_GENIEv3_sumtp} for the \texttt{GENIEv3} model. Due to a unphysical modelling of the multi-pion production channel in the \texttt{GENIEv3} model, only the QE and 2p2h contributions are reweighted, while the pion/hadron production contribution is kept identical to the reference model \texttt{MnvTune4.4.1}.

%%%%%%%%%%%%%%% NEUT-SF %%%%%%%%%%%%%%%%%%%%%%%%%%%%%%%%

%%%%%%%%%%%%%%% NuWro-LFG %%%%%%%%%%%%%%%%%%%%%%%%%%%%%%%%

%%%%%%%%%%%%%%% NuWro-LFG %%%%%%%%%%%%%%%%%%%%%%%%%%%%%%%%

%%%%%%%%%%%%%%% GENIEv3 %%%%%%%%%%%%%%%%%%%%%%%%%%%%%%%%

\clearpage